%% file: main.tex
\pdfoutput=1
\documentclass[journal]{IEEEtran}
\usepackage[utf8]{inputenc}
\usepackage[T1]{fontenc}
\usepackage{amsmath,amssymb,amsfonts}
\usepackage{graphicx}
\usepackage{booktabs}
\usepackage{multirow}
\usepackage{array}
\usepackage{tabularx}
\usepackage{xcolor}
\usepackage[hidelinks]{hyperref}
\usepackage{cite}
\usepackage{url}
\usepackage{microtype}
\input{numbers_rev.tex}

\input{numbers_tip.tex}
\input{figs/fig_refinements_key.tex}

\newcommand{\ca}{C\&A}
\newcommand{\drift}{\mathrm{D}}

\begin{document}
\bstctlcite{IEEEtranBSTCTL}

\title{Albedo Drift: Measuring the Illumination-Dependent Error of Intrinsic Image Decomposition}

\author{Jihwan~Woo%
\thanks{J. Woo is a Senior Specialist Solutions Architect, AI/ML, with Amazon Web Services. E-mail: jihwan.woo@gmail.com.}%
\thanks{A preliminary version of the drift measurement was submitted to a machine-learning conference in 2026 and is under review at the time of posting. This article is a complete rewrite with a new ground-truth analysis, a downstream evaluation and an exposure study, and it withdraws the earlier version's claim that the invariance can be improved at no cost in fidelity.}}

\markboth{Preprint, November 2026}{Woo: Albedo Drift}

\maketitle

\begin{abstract}
Intrinsic image decomposition recovers from one photograph an albedo layer that should not change when the lighting changes. The standard benchmarks never test this: they score one photograph under one light. We measure the property directly. Albedo drift is the variation of the predicted albedo across photographs of one scene under different lighting, divided by that of the photographs: a correct albedo scores zero, copying the photograph scores one. On 30 multi-illumination scenes the learned models keep \TMLDriftBaseMin--\TMLDriftBaseMax{} of the photograph's change; on indoor renders with ground truth under two lightings, \TORDriftBaseMin--\TORDriftBaseMax; on time-lapse video, more. Drift is only a necessary condition, since a flat albedo is perfectly invariant, so we add a contrast-matched form and a ground-truth anchor. The anchor shows that the drift numerator is a ground-truth-free surrogate of the illumination-dependent component of the albedo error, which for the published models is \TORFdepBaseMin--\TORFdepBaseMax{} percent of that error on the renders and \TMITFdepBaseMin--\TMITFdepBaseMax{} percent on lab objects under ten directional lights. Across operating points including adapters and fine-tunes that apply a cross-illumination consistency loss to a trained model, lower drift accompanies higher ground-truth error (rank correlation \TORAtwoAllFive{} on \TORAtwoAllN{} and \TMITAtwoAllFive{} on \TMITAtwoAllN{} points): the objective lowers the ratio by flattening the albedo, raising ground-truth and relighting error up to four-fold. The contrast-matched form is positively correlated (\TORAtwoAllSix{} and \TMITAtwoAllSix). Invariance should be measured with contrast and fidelity, never optimized alone.
\end{abstract}


\IEEEpeerreviewmaketitle

\section{Introduction}
\IEEEPARstart{A}{n} intrinsic image decomposition splits a photograph $I$ into an albedo $R$ and a shading $S$ with $I = R \odot S$ \cite{barrow78intrinsic}. The definition of albedo is that it belongs to the surface: when the light changes, $S$ changes and $R$ does not. Yet the two evaluation protocols that the field has used for a decade score one photograph under one light. The Weighted Human Disagreement Rate on Intrinsic Images in the Wild \cite{bell14iiw} asks whether pairs of points inside one image have the right relative reflectance; the scale-invariant error on MIT Intrinsics \cite{grosse09mit} compares one prediction with the ground-truth reflectance of one object under one light. Neither protocol separately diagnoses an albedo that is plausible in every single image and different in every one: the ground-truth error penalizes it as part of a total, the human judgments not at all (Fig.~\ref{fig:qual}).

This article measures that failure, which we call \emph{albedo drift}. Given photographs of one scene under several lightings, drift is the cross-illumination variation of the predicted albedo divided by that of the photographs, after each image is normalized by its own median to remove the scale ambiguity of the decomposition. A correct albedo scores $0$; returning the photograph scores exactly $1$. The ratio needs no ground truth and applies to any method, so real photographs suffice. On the test split of the Multi-Illumination dataset \cite{murmann19multilum}, the best learned model keeps \BestExistingML{} of the photograph's change in its albedo and the diffusion models keep up to \MGMLMed; on synthetic indoor renders with ground truth under two lighting configurations \cite{li21openrooms} the learned models keep \TkoORDrift--\TcaoORDrift; on time-lapse video \cite{li18bigtime} the normalized leak is two to three times larger. Similar IIW human-disagreement rates do not imply similar drift: two models with the same WHDR differ two-fold in multilum drift.

A low drift ratio, however, is not evidence of a correct albedo. A constant image has zero drift, and an albedo whose spatial contrast has been halved keeps less of the light because it keeps less of everything. We therefore treat drift as a \emph{necessary condition} and give it two companions: a contrast-matched form, in which every frame is standardized to unit spatial contrast before the cross-illumination variation is taken, and a ground-truth anchor where ground truth exists. The anchor is more than a check: the scale-aligned error of a prediction splits orthogonally into a part shared by all lightings and a part that changes with the light, and the second is built from the per-pixel residual that the drift numerator measures without ground truth (median instead of least-squares scaling, display units, absolute instead of squared aggregation). Drift is therefore a ground-truth-free surrogate of one component of correctness, the \emph{illumination-dependent error}. On the two ground-truth datasets this component is a modest part of the total for the published base models: \TORFdepBaseMin--\TORFdepBaseMax\% of the error on the indoor renders, where the two lightings share most of their shading structure, and \TMITFdepBaseMin--\TMITFdepBaseMax\% on lab objects under ten directional lights. The rest is a residual shared across the evaluated lightings, invisible to differences within that set. The component is worth tracking nonetheless: it is the part of the error accessible without ground truth through cross-lighting differences, it can contribute to the cost of re-using an albedo under a different light (Section~\ref{sec:res_down}), and optimizing it in isolation is harmful.

The consequence for method development is the article's central result. We evaluated \TORAtwoAllN{} operating points on the renders (\TMITAtwoAllN{} on the lab objects): six published models, eleven refinements of two of them that apply a cross-illumination consistency objective of the kind used since \cite{li18bigtime} to the trained model, as a residual adapter on the albedo or by fine-tuning the albedo network, and two trivial smoothing filters. Across these points lower drift accompanies \emph{higher} ground-truth error (rank correlation \TORAtwoAllFive{} on the renders, \TMITAtwoAllFive{} on the lab objects), because every refinement lowers the ratio mostly by flattening the albedo; the contrast-matched form is positively correlated (\TORAtwoAllSix{} and \TMITAtwoAllSix). Among the eleven \ca{}-derived points (nine trained refinements, two filters), the more a point lowers the drift ratio the more it raises the ground-truth error (rank correlation \TORAthree{} and \TMITAthree) and the relighting error of a downstream re-use of the albedo (up to \TMITdEcrossMax\%). Invariance can be gamed when optimized in isolation, as any necessary condition can; this argues for measuring it with contrast and fidelity, not for leaving it unmeasured.

The contributions are: (1) a drift protocol with matched encoding, identity controls and a contrast-matched form, measured on eight methods and five datasets; (2) an exact decomposition of the ground-truth error into a shared and an illumination-dependent part, the relation of the drift numerator to the second, measured on two ground-truth datasets; (3) the finding that a consistency objective applied to a trained model, as an adapter or a fine-tune, lowers drift by flattening and thereby worsens ground-truth and relighting error, so that the plain ratio ranks operating points in the wrong order; (4) a downstream relighting measurement: the excess cost of re-use was near zero on the renders and \TMITBoneBaseMin--\TMITBoneBaseMax\% on the lab objects, the dataset with the larger illumination-dependent fraction; (5) a protocol study showing that the drift ratio depends on the exposure policy applied before 8-bit encoding by up to \TXDriftRefMin\% while ground-truth error changes by a few percent at most, so the policy must be fixed and reported. All experiments were pre-registered with expected ranges and failure conditions; three hypotheses failed and are reported as such (Section~\ref{sec:failed}).

\input{sections/related}

\section{Albedo Drift}
\label{sec:method}
This section constructs the measurement from the requirements of Table~\ref{tab:gaps}. The drift ratio (Section~\ref{sec:drift}) answers the Single-Light Problem by comparing the albedo of one scene under two lights (R1) after a scale normalization (R2); the encoding discipline and the identity control (Section~\ref{sec:encoding}) answer the Encoding Problem (R4); the contrast-matched form (Section~\ref{sec:cm}) answers the Open-Contrast Problem (R3); the ground-truth decomposition (Section~\ref{sec:decomp}) answers the Undecomposed-Error Problem and relates the drift numerator to one of its two parts (R5); and the relighting measurement (Section~\ref{sec:downstream}) answers the Re-use Problem (R5$'$). Each quantity is defined so that a trivial method has a known value, which Section~\ref{sec:controls} turns into controls.

\subsection{Image formation and the quantity measured}
For a scene photographed under lightings $\ell = 1, \dots, L$ with a fixed camera, $I_\ell = R \odot S_\ell$ holds per pixel and per channel for a Lambertian surface, with the albedo $R$ shared by all $\ell$, provided the camera's channel responses are narrow enough for the spectral integral to factor; for broadband channels under a spectrally changing light the factorization is an assumption, $R$ a canonical RGB reflectance and $S_\ell = I_\ell / R$ an effective shading absorbing the spectral interaction (the operational definition of Section~\ref{sec:downstream}); the renders used for ground truth factor by construction. A decomposition method returns $A_\ell = f(I_\ell)$. Because $I = (cR)\odot(S/c)$ for any positive $c$, albedo is defined only up to a global (for colour, per-channel) scale, so every comparison below first normalizes each image by its own median: the primary, luminance-normalized form removes one scalar per frame and keeps chromatic drift measurable (so ``a correct albedo scores zero'' presupposes a consistent colour convention across lightings); the per-channel variant removes the diagonal gain too. Non-Lambertian pixels violate the model; the primary multilum measurement uses the full frame with a highlight-rejection variant beside it, and the other datasets use the masks of Section~\ref{sec:setup}.

\subsection{The drift ratio}
\label{sec:drift}
Let $\tilde{X} = X / \mathrm{median}(X)$ denote an image divided by the median of its channel mean over a mask $\Omega$ (the luminance variant; the per-channel variant divides each channel by its own median). The drift ratio of a method on a scene is
\begin{equation}
\drift = \frac{\overline{\sigma_\ell(\tilde{A}_\ell)}}{\overline{\sigma_\ell(\tilde{I}_\ell)}},
\label{eq:drift}
\end{equation}
where $\sigma_\ell(\cdot)$ is the standard deviation across lightings at a pixel and the bar is the mean over $\Omega$. The denominator is the normalized change of the photographs, the numerator that of the predicted albedos. A correct albedo gives $\drift = 0$ up to the median normalization; the identity $A_\ell = I_\ell$, or the photograph re-scaled, gives exactly $1$. $\drift$ is not bounded above: a value above $1$ means that the albedo changes more than the photograph. Its absolute value depends on how much a dataset's photographs change, so values are comparable within a dataset and across datasets only in rank. We report the median of $\drift$ over scenes with a scene-bootstrap 95\% confidence interval, and means where a comparison to earlier work requires them.

\subsection{Encoding discipline}
\label{sec:encoding}
Photographs are stored in a display-referred encoding (sRGB-like; Section~\ref{sec:setup}); every learned model evaluated here returns a \emph{linear} albedo. Comparing the two directly is a mixed convention under which a no-op scores far above one: the linearized photograph $I^{2.2}$ fed into \eqref{eq:drift} as the ``albedo'' gives \IDLINMLMed{} on the multi-illumination scenes and \IDLINBTMed{} on time-lapse video, because the gamma curve itself changes the normalized contrast. The protocol therefore re-encodes every linear output with the inverse of the model's own linearization ($A^{1/2.2}$) before \eqref{eq:drift}, the stored photograph being the reference. Each model's input convention is read from its code, not its documentation: the diffusion model of \cite{kocsis24iid} linearizes its input with exponent $2.2$, RGB$\to$X \cite{zeng24rgbx} applies a $1/2.2$ gamma to its output, which we invert, and NIID-Net \cite{li20niid} consumes the encoded image and returns linear reflectance at $320\times240$. Two controls run under every convention: the identity (the stored photograph returned as the albedo) must score exactly $1$; the identity in linear units (the linearized photograph, not re-encoded) exposes a mixed convention by scoring far above $1$.

\subsection{Contrast-matched drift}
\label{sec:cm}
A constant image has $\drift = 0$, and an albedo with half the spatial contrast of another keeps roughly half as much of the light. The contrast-matched ratio removes this degree of freedom by standardizing every frame over the mask to zero mean and unit spatial standard deviation before the cross-illumination variation is taken,
\begin{equation}
\drift_z = \frac{\overline{\sigma_\ell(z(\tilde{A}_\ell))}}{\overline{\sigma_\ell(z(\tilde{I}_\ell))}}, \qquad z(X) = \frac{X - \mu_\Omega(X)}{\sigma_\Omega(X)}.
\label{eq:cm}
\end{equation}
The identity still scores exactly $1$. The standardization follows division by the median luminance, so no absolute constant enters and a rescaled prediction receives the same $\drift_z$ to floating-point precision; a frame whose standard deviation divided by its median is below $10^{-3}$, or whose median luminance is below $10^{-6}$, has no defined $z$ and its $\drift_z$ is undefined (an exactly constant frame would otherwise standardize to zeros and score $\drift_z = 0$). Every scoring path (the plain and contrast-matched scorers and the relighting analysis' recomputation of the plain ratio) uses one implementation of these rules: photographs whose normalized variation is below $10^{-6}$ (identical frames, frames differing by a global gain only or, for $\drift_z$, by an offset only) define no ratio; a prediction that does not change between two different photographs keeps $\drift = 0$; and a gain sweep of a copied frame from $1$ to $\TgainSweepMinGain$ returns $\drift$ and $\drift_z$ within $\TgainSweepMaxDev$ of $1$ down to a gain of $\TgainSweepMinDefined$ and undefined from $\TgainSweepMaxUndefined$ on, never a favourable finite score. The archived values were computed with absolute floors ($10^{-6}$ on the median, $10^{-8}$ on the denominator) that, by the audit below, never activated, so none changes. The rules were audited over every cached photograph and prediction frame of the five acquisition conditions with cross-lighting stacks (multilum, its RAW-derived re-renders, MIT Intrinsics, OpenRooms, BigTime; \TminFrameN{} frames): the smallest normalized deviation is \TminFrameStd, the smallest median luminance \TminFrameMedian, and the smallest denominator \TminDen{} for $\drift$ and \TminDenZ{} for $\drift_z$, so no evaluated quantity is near a threshold; a near-constant frame that passes them has its small residual variation amplified. What the form removes exactly is a positive affine change $aX + b$ of the display-encoded prediction; a contraction of the \emph{linear} albedo towards a constant is not affine after encoding and is not cancelled: $\drift_z$ controls for global contrast changes in its evaluation domain, not for every flattening. In \eqref{eq:cm}, $\mu_\Omega$ and $\sigma_\Omega$ are the mean and standard deviation over $\Omega$ of the luminance (channel mean), applied to all three channels, and $\sigma_\ell$ is averaged over channels as in \eqref{eq:drift}; the per-channel variant standardizes each channel by its own statistics, and forgives a change of colour balance between the lights that the luminance variant keeps measurable. We report $\drift$, $\drift_z$ and the albedo's spatial contrast relative to the photograph ($\sigma_\Omega(\tilde{A})/\sigma_\Omega(\tilde{I})$, averaged over frames) together; the three expose contrast loss where it explains a low ratio and flag exact degeneracy, without certifying material recovery.

\subsection{Ground-truth decomposition of the error}
\label{sec:decomp}
Where a ground-truth albedo $R$ exists, let $\alpha_\ell$ be the least-squares scale that aligns $A_\ell$ with $R$ over $\Omega$ (the alignment of the scale-invariant error of \cite{grosse09mit}) and $\varepsilon_\ell = \alpha_\ell A_\ell - R$ the aligned residual, whose mean square over $\Omega$ is the scale-invariant error (si-MSE) $e_\ell$ of lighting $\ell$. Define the shared and the illumination-dependent residuals
\begin{equation}
\bar{\varepsilon} = \tfrac{1}{L}\textstyle\sum_\ell \varepsilon_\ell, \qquad \delta_\ell = \varepsilon_\ell - \bar{\varepsilon}.
\label{eq:split}
\end{equation}
Because $\sum_\ell \delta_\ell = 0$, the cross terms vanish and
\begin{equation}
\tfrac{1}{L}\textstyle\sum_\ell \|\varepsilon_\ell\|^2 = \|\bar{\varepsilon}\|^2 + \tfrac{1}{L}\textstyle\sum_\ell \|\delta_\ell\|^2,
\label{eq:identity}
\end{equation}
an exact orthogonal decomposition of the mean ground-truth error into the sample-mean residual shared within the evaluated lighting set and a part that changes with the light; a different set of lightings can change both parts. We call
\begin{equation}
f_{\mathrm{dep}} = \frac{\frac{1}{L}\sum_\ell \|\delta_\ell\|^2}{\|\bar{\varepsilon}\|^2 + \frac{1}{L}\sum_\ell \|\delta_\ell\|^2}
\label{eq:fdep}
\end{equation}
the illumination-dependent fraction of the error. The ground truth $R$ cancels in $\delta_\ell$---for two lightings $\delta = (\alpha_a A_a - \alpha_b A_b)/2$---but the alignment scales $\alpha_\ell$ still require it. The numerator of \eqref{eq:drift} is built from the same per-pixel residual with two substitutions that need no ground truth---median normalization for the least-squares scales, display encoding for linear units---and aggregates it by its mean absolute value ($\sigma_\ell$ at a pixel is $|\tilde A_a - \tilde A_b|/2$ for two lightings) where \eqref{eq:identity} uses the mean square. It is therefore a ground-truth-free \emph{surrogate} of the illumination-dependent term, not the term itself; empirically the two rank the views alike: the Spearman correlation between the numerator of \eqref{eq:drift} and the root-mean-square of $\delta$ relative to that of the ground-truth albedo, $\mathrm{rms}(\delta)/\mathrm{rms}(R)$, is \TORRhoNumMin--\TORRhoNumMax{} per method on the indoor renders and \TMITRhoNumMin--\TMITRhoNumMax{} on the lab objects (the normalization by $\mathrm{rms}(R)$ matches the scale-free numerator; the unnormalized $\|\delta\|$ is a different target). \eqref{eq:identity} is exact; \eqref{eq:fdep} is the second term's share of the total error, whereas the association validated above is with the term's normalized magnitude.

\subsection{Downstream re-use of the albedo}
\label{sec:downstream}
The application that motivates invariance is re-use: an albedo estimated under one lighting is combined with a different shading, which ground truth lets us test without a shading estimator. For lightings $a \neq b$ let $S_b = I_b / R$ be the ground-truth shading of $b$ (the identity $I = R\odot S$ applied to the stored photograph, so specular and interreflected light are in $S_b$). The cross-lighting relight $\hat{I}_b^{(a)} = \alpha\, A_a \odot S_b$ and the same-lighting relight $\hat{I}_b^{(b)} = \alpha\, A_b \odot S_b$, each with a least-squares $\alpha$, are compared with $I_b$ by the relative squared error $E = \sum_\Omega(\hat{I}-I_b)^2 / \sum_\Omega I_b^2$ in linear RGB. Since $S_b = I_b / R$, $E = \sum_\Omega S_b^2 (\alpha A_a - R)^2 / \sum_\Omega S_b^2 R^2$: an albedo error weighted by the target shading with a per-pair scale, unlike the unweighted decomposition of \eqref{eq:identity}. Averaged over all ordered pairs, $\Delta E = E_{\mathrm{cross}} - E_{\mathrm{within}}$ is the excess error caused by re-using an albedo across lightings. The ground-truth albedo gives $E = 0$ exactly; a constant albedo gives $\Delta E = 0$ and an $E$ that measures how much the albedo's spatial structure matters at all.

\subsection{Controls and record}
\label{sec:controls}
Every result file carries the following controls. Three must hold exactly: the identity scores $\drift = \drift_z = 1$; the ground-truth albedo returned as the prediction scores $\drift = 0$, zero scale-invariant error and $E = 0$; the photograph returned as the albedo is the display-encoded image for the drift ratios (scoring exactly $1$) and the linear image for every ground-truth and relighting quantity; a constant albedo gives $\Delta E = 0$. One is a tested diagnostic: the drift between the albedo of one view and that of a \emph{different} view of the same scene (mismatched pairs) must have a rank correlation below $0.15$ in magnitude with the view's lighting sensitivity; it does for every method. The record of every number is a cache of the predicted albedos in 16-bit floating point, from which the archived statistics are recomputed so that later analyses reproduce them to the last digit; the main statistics were recomputed through a second code path. Two methods were run on two GPU hosts; their per-view drift ratios agree to \TORxhostMax{} (GPU nondeterminism).

\section{Data, Models and Protocol}
\label{sec:setup}

\subsection{Datasets}
\textbf{Multi-illumination (multilum).} The 30 test scenes of \cite{murmann19multilum}, each photographed under 25 directional flashes with a fixed camera; the stored photographs are the dataset's tone-mapped 8-bit renderings of its HDR captures, at 768 pixels. The primary measurement uses every pixel; the ``diffuse-only'' variant is a highlight-rejection heuristic on the channel-mean intensity (pixels whose maximum over the 25 frames is $\geq 0.97$ or exceeds the per-pixel median over frames by $\geq 0.5$, grown by seven binary dilations, are removed) applied to the aggregation of \eqref{eq:drift} after full-frame median normalization. \textbf{RAW-derived renders.} The same 30 scenes re-rendered from the RAW brackets through a fixed pipeline (merge, fixed white balance, one exposure per scene, sRGB curve; supplement), a control for the camera's tone mapping. \textbf{MIT Intrinsics.} The 20 objects of \cite{grosse09mit}, each under 10 directional lights with a ground-truth reflectance and an object mask, which is the only mask applied. The files are linear: the objects were captured with a linear camera \cite{grosse09mit}, and in the distributed files the original image equals diffuse plus specular to within quantization (median absolute difference \TMITaddMed{} in 8-bit units over \TMITaddN{} object masks, 99th percentile \TMITaddPnn), which a power-law or sRGB encoding would break. They are encoded with $x^{1/2.2}$ before inference, so that every model receives the convention it was trained on, and the ground truth stays linear. \textbf{BigTime.} 30 time-lapse sequences of \cite{li18bigtime} without ground truth, 12 frames each under the dataset's validity and sky masks (selection rule in the supplementary material); scene changes between frames are not separated from the lighting change. \textbf{OpenRooms.} The test split of OpenRooms \cite{li21openrooms} (\TORscenes{} scenes) provides, per camera view, renders under two lighting configurations (\texttt{main} and \texttt{mainDiffLight}) that share geometry and materials and differ in their light sources (the dataset's documentation), a ground-truth diffuse albedo, and light-source masks; \TORnSel{} views were sampled at random (seed 0). The renders are linear radiance; both frames of a view are tone-mapped with one shared exposure (the median radiance of the emitter-free pixels of the two frames mapped to 0.18), clipped, encoded with $x^{1/2.2}$ and quantized to 8 bits. This emulates a fixed camera with a global exposure only (no white balance, noise, demosaicing or tone curve beyond the power law); the exponent $1/2.2$ rather than the sRGB curve lets the identity-in-linear control recover the exposed radiance on unclipped pixels, up to 8-bit quantization. The mask excludes emitters in either lighting, pixels whose ground-truth albedo is below $0.02$ in any channel, and pixels clipped in either frame. \TORskipBlack{} black renders and \TORskipMask{} views with less than 5\% valid pixels are excluded (rules and clipping statistics in the supplementary material); \TORn{} views are analysed. The ground-truth albedo files are gamma-encoded and were linearized with the exponent $2.2$, verified against the dataset's own shading (supplementary material). \textbf{IIW.} The Weighted Human Disagreement Rate on the full test set of \cite{bell14iiw} where available, else on a fixed 300-image subset.

\subsection{Methods}
\emph{Regression:} \ca{} v2 \cite{careaga24colorful} (also its 2023 weights \cite{careaga23intrinsic}) and NIID-Net \cite{li20niid}. \emph{Diffusion:} Marigold-IID \cite{ke25marigold} with one sample per frame (a ten-sample ensemble is a control), the intrinsic image diffusion of \cite{kocsis24iid}, and RGB$\to$X \cite{zeng24rgbx} with its albedo prompt and 50 DDIM steps. Noise coupling across the frames of a scene differs: the Marigold runs advance one generator (seed 0) through the frames (a different noise realization per frame), whereas the Kocsis and RGB$\to$X runs re-seed before every frame (common initial noise); Section~\ref{sec:res_drift} measures the difference, and under the first policy the varying component of the error decomposition includes sampling variation. \emph{Classical control:} a log-domain single-scale centre/surround operator of the Retinex family ($\sigma = 32$ pixels on the logarithm of the display-encoded photograph; definition in the supplementary material), labelled ``Retinex (log-domain)'' because \cite{jobson97ssr} forms the surround on the intensity. \emph{References:} the identity, the ground-truth albedo where it exists, and two trivial refinements of the \ca{} albedo, a Gaussian blur ($\sigma = 4$ pixels) and a bilateral filter ($d = 9$, range $0.1$, space $7$ pixels), both on the linear albedo. The models' training data, as stated by their authors, are tabulated in the supplementary material: both \ca{} models used albedo derived from the multi-illumination dataset; none lists OpenRooms, MIT Intrinsics or IIW test images, though four were trained on synthetic indoor renders of OpenRooms' kind, and whether NIID-Net trained on IIW is not stated in its code release.

\subsection{Consistency objectives as probes}
\label{sec:probes}
To test what optimizing the invariance does, we trained two kinds of refinement with a cross-illumination consistency loss on the 60 training scenes of multilum: a residual U-Net \emph{adapter} on the \ca{} albedo, conditioned on the image, trained on pairs $(a, b)$ of lightings with the objective
\begin{equation}
\mathcal{L} = \|\tilde{A}_a - \tilde{A}_b\|_1 + \tfrac{\lambda}{2}\big(\|A_a - A^{\mathrm{base}}_a\|_1 + \|A_b - A^{\mathrm{base}}_b\|_1\big),
\label{eq:probe}
\end{equation}
in which $A$ is the refined linear albedo, $\tilde A$ its median-normalized form (as in \eqref{eq:drift}) and $A^{\mathrm{base}}$ the base model's albedo: a consistency term on the normalized albedos plus an anchor of weight $\lambda$ in linear units ($\lambda = 1$ for 20k steps, $\lambda \in \{0.3, 3\}$ and an albedo-only variant for 10k, and one re-run of $\lambda = 1$ with the same seed, differing only by GPU nondeterminism, evaluated at 10k and 20k steps), and a \emph{fine-tune} of the \ca{} albedo network with the same objective ($\lambda = 1$ for 10k steps, $\lambda \in \{3, 10\}$ for 5k). Batch, crop, learning rates and every checkpoint's stored configuration are tabulated in the supplementary material. The anchor is contrast-sensitive, but the consistency term outweighs it wherever a lower-contrast albedo reduces the cross-lighting difference by more than it costs in anchor distance; what the trade-off produces is measured, not assumed. The same adapter recipe with $\lambda = 1$ was trained on NIID-Net as a second base model. These models are probes of the objective, not proposed methods, and the article's claims about them are negative; every operating point reported is a base model or one of these multilum-trained probes (refinements trained on other data belong to a separate study).

\subsection{Pre-registration}
Every experiment on the ground-truth datasets was written down before it ran, with the quantity, the expected range, a failure condition and a ``too strong'' threshold that triggers a search for a defect before any result is reported. Three hypotheses failed their conditions and one expectation was contradicted in the direction of a stronger effect (Section~\ref{sec:failed}); the expected ranges and outcomes are listed in the supplementary material.

\textbf{Statistical reporting.} The resampling unit is the scene (multilum, BigTime), the object (MIT) or the scene with all its views (OpenRooms); confidence intervals are percentile bootstrap intervals over that unit (10,000 resamples for multilum, BigTime and the MIT drift ratios, 2,000 for the OpenRooms and MIT ground-truth analyses; seed 0). Paired changes between a refinement and its base are tested with a two-sided sign-flip permutation test of the \emph{mean} per-unit difference (10,000 flips; $p$ is the fraction of flips at least as extreme as the observation; ``$p < 10^{-4}$'': no flip reached it, corrected estimate $1/10{,}001$), even where nearby percentage changes are of medians; ``worse in $x$\% of views'' is the fraction of units with a larger error. Correlations are Spearman's; their intervals, where given, are cluster bootstraps over scenes.

\section{Results}
\label{sec:results}

\subsection{Drift across datasets}
\label{sec:res_drift}
Table~\ref{tab:main} reports the base models. On multilum, the best-case in-distribution setting for \ca{} v2 (its training data include albedo derived from this dataset \cite{careaga23intrinsic,careaga24colorful}; the authors report 1,000 of its scenes and do not state whether the 30 test scenes were excluded), the model keeps \CAMLMed{} [\CAMLLo, \CAMLHi] of the photograph's normalized change (at least \CAMLMin{} in every scene, at least a fifth in \CAMLnAboveFifth{} of 30). Diffuse pixels alone give \CAMLDif, the RAW-derived renders \CAHDRMed{} [\CAHDRLo, \CAHDRHi] (Marigold-IID \MGHDRMed, RGB$\to$X \RXHDRMed; the other models were not run on this control), and per-channel normalization \CAMLPc: the leak persists under the highlight-rejection mask, under the dataset's tone mapping and under per-channel normalization, which forgives any global colour cast. The diffusion models keep more (Marigold-IID \MGMLMed{} [\MGMLLo, \MGMLHi], RGB$\to$X \RGBXMLMed); sampling noise contributes to but does not create Marigold-IID's drift: five seeds on four scenes put its same-image variation at \MGseedOverIllum{} (\MGseedOverIllumMin--\MGseedOverIllumMax) of its cross-illumination variation; re-running those scenes with common initial noise across frames (the coupling of the other two diffusion models) changes its drift ratio by \TmgNoiseRelMed\% (median of the per-scene changes, \TmgNoiseRelMin\% to \TmgNoiseRelMax\%; the medians over scenes move \TmgNoiseAdvMed{} $\to$ \TmgNoiseFixMed) and its contrast-matched drift by \TmgNoiseCmRelMed\%, so the archived values stand with that qualification; a ten-sample ensemble lowers its median only to \MGensMLMed.

\begin{table*}[t]
\centering
\caption{Base models on five datasets. Drift: median of \eqref{eq:drift} (matched encoding, luminance) over scenes (multilum), objects (MIT), sequences (BigTime) or views (OpenRooms), with a 95\% cluster-bootstrap interval over scenes where computed; $\drift_z$: contrast-matched \eqref{eq:cm}; contrast: albedo spatial contrast relative to the photograph; WHDR: full IIW ($\ddagger$: 300-image subset); si-MSE: median over views (OpenRooms) or objects (MIT) of the mean over their lightings of the scale-invariant error against ground truth (same nesting in Table~\ref{tab:ops} and Fig.~\ref{fig:inv_err}). OpenRooms: \TORn{} views under two lightings; MIT: 20 objects under 10 lights; BigTime: 30 time-lapse sequences. ---: not run (BigTime for \ca{} 2023 and Retinex) or not computed (Retinex contrast). Ground-truth error of the identity and of Retinex is computed in linear units (linear photograph, linearized Retinex output), the reference of the relighting measurement; the display-encoded versions used by the drift columns give \TidMITSim{}/\TidORSim{} (identity) and \TrxMITSim{}/\TrxORSim{} (Retinex).}
\label{tab:main}
\footnotesize
\begin{tabular}{l ccc c cc cc ccc c}
\toprule
 & \multicolumn{3}{c}{multilum (30 scenes)} & IIW & \multicolumn{2}{c}{MIT Intrinsics} & \multicolumn{3}{c}{OpenRooms} & BigTime \\
\cmidrule(lr){2-4} \cmidrule(lr){5-5} \cmidrule(lr){6-7} \cmidrule(lr){8-10} \cmidrule(lr){11-11}
Method & Drift [CI] & $\drift_z$ & contrast & WHDR & Drift & si-MSE & Drift & $\drift_z$ & si-MSE & Drift \\
\midrule
\ca{} v2 \cite{careaga24colorful} & \CAMLMed{} [\CAMLLo, \CAMLHi] & \CAMLCm & \CAMLCon & \CAWhdr & \TcaMITDrift & \TcaMITSim & \TcaORDrift & \TcaORCm & \TcaORSim & \CABTMed \\
\ca{} 2023 \cite{careaga23intrinsic} & \CAoMLMed & \CAoMLCm & \CAoMLCon & \CAoWhdrThree$^\ddagger$ & \TcaoMITDrift & \TcaoMITSim & \TcaoORDrift & \TcaoORCm & \TcaoORSim & --- \\
NIID-Net \cite{li20niid} & \NIIDMLMed{} [\NIIDMLLo, \NIIDMLHi] & \NIIDMLCm & \NIIDMLCon & \NIIDWhdrThree$^\ddagger$ & \TniidMITDrift & \TniidMITSim & \TniidORDrift & \TniidORCm & \TniidORSim & \NIIDBTMed \\
Kocsis et al. \cite{kocsis24iid} & \KOMLMed{} [\KOMLLo, \KOMLHi] & \KOMLCm & \KOMLCon & \KOWhdrThree$^\ddagger$ & \TkoMITDrift & \TkoMITSim & \TkoORDrift & \TkoORCm & \TkoORSim & \KOBTMed \\
RGB$\to$X \cite{zeng24rgbx} & \RGBXMLMed{} [\RGBXMLLo, \RGBXMLHi] & \RGBXMLCm & \RGBXMLCon & \RGBXWhdrThree$^\ddagger$ & \TrgbxMITDrift & \TrgbxMITSim & \TrgbxORDrift & \TrgbxORCm & \TrgbxORSim & \RGBXBTMed \\
Marigold-IID \cite{ke25marigold} & \MGMLMed{} [\MGMLLo, \MGMLHi] & \MGMLCm & \MGMLCon & \MGWhdr & \TmgMITDrift & \TmgMITSim & \TmgORDrift & \TmgORCm & \TmgORSim & \MGBTMed \\
Retinex (log-domain) & \RXMLMed & \RXMLCm & --- & \RXWhdr & \TrxMITDrift & \TrxlMITSim & \TrxORDrift & \TrxORCm & \TrxlORSim & --- \\
Identity (photograph) & 1.000 & 1.000 & 1.00 & \IDWhdr & 1.000 & \TidlMITSim & 1.000 & 1.000 & \TidlORSim & 1.000 \\
Ground-truth albedo & --- & --- & --- & --- & 0.000 & 0.0000 & 0.000 & --- & 0.0000 & --- \\
\bottomrule
\end{tabular}
\end{table*}

\textbf{Similar IIW WHDR does not imply similar drift.} \ca{} v2 and Marigold-IID have nearly the same WHDR on the full IIW test set (\CAWhdr{} and \MGWhdr) and a two-fold difference in drift (\CAMLMed{} and \MGMLMed); the model of \cite{kocsis24iid} matches \ca{}'s drift (\KOMLMed) at a worse WHDR (\KOWhdrThree), and NIID-Net scores \NIIDMLMed{} at \NIIDWhdrThree. Contrast-matched, the ranking changes again: Marigold-IID (\MGMLCm, albedo contrast \MGMLCon{} of the photograph's) is as stable as \ca{} (\CAMLCm{} at contrast \CAMLCon), while Kocsis et al.\ (\KOMLCm) and NIID-Net (\NIIDMLCm) are the least stable per unit of contrast: their low ratios are partly low contrast (\KOMLCon{} and \NIIDMLCon).

\textbf{Normalized drift is higher on the other datasets.} Absolute ratios are comparable within a dataset only (Section~\ref{sec:drift}). On BigTime, \ca{}'s median drift is \CABTMed{} [\CABTLo, \CABTHi], three times its multilum value, with \CABTFrac{} of sequences above $1$; the diffusion models rise two- to three-fold (Kocsis et al.\ \KOMLMed{}$\to$\KOBTMed, RGB$\to$X \RGBXMLMed$\to$\RGBXBTMed, Marigold-IID \MGMLMed$\to$\MGBTMed) and NIID-Net by half; content, lighting statistics and the denominator all differ between datasets, so this describes rather than attributes to training-domain shift. On the OpenRooms pairs, lit by different light sources, the learned models keep \TkoORDrift--\TcaoORDrift{} of the photograph's change (Table~\ref{tab:main}); contrast-matched, \TrgbxORCm--\TcaoORCm: per unit of contrast their albedo moves almost as much as the photograph. The log-domain Retinex control is less invariant than \ca{} on multilum (\RXMLMed) and far worse on the lab objects (\RXMITeMed); for this operator the ranking against the learned methods is domain-dependent.

\subsection{Invariance versus correctness}
\label{sec:res_corr}
\begin{figure*}[t]
\centering
\includegraphics[width=0.65\linewidth]{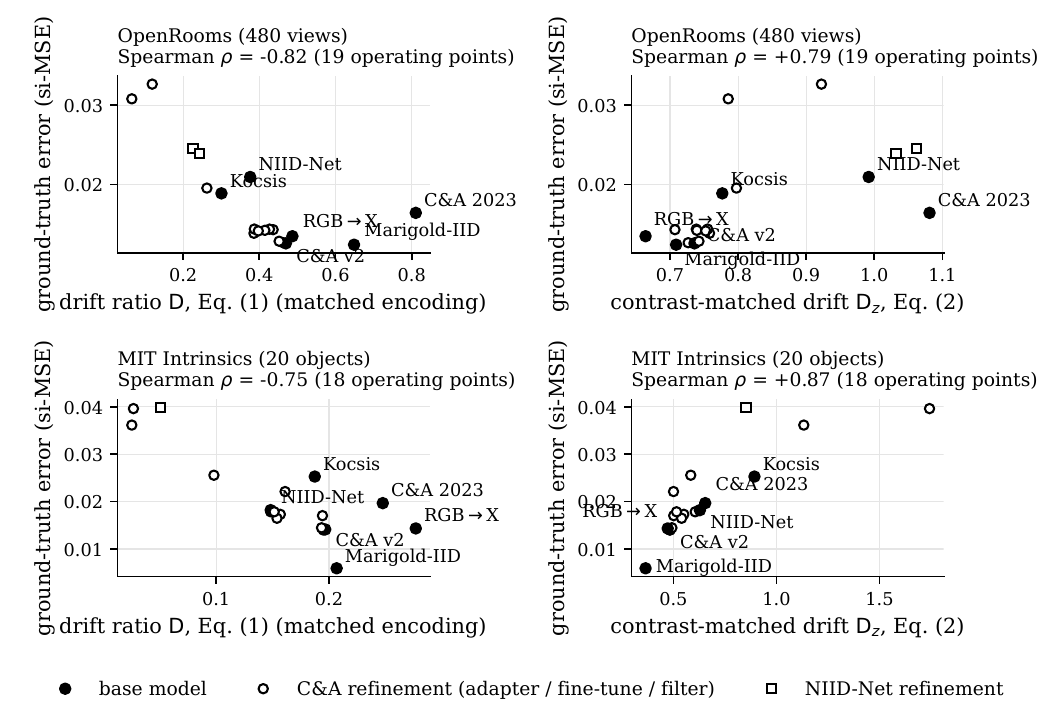}
\caption{Invariance against correctness. Each point is one operating point (filled: base model; hollow circle: refinement of \ca{} v2 by an adapter, a fine-tune or a filter; hollow square: refinement of NIID-Net); $x$ is the median drift over views (OpenRooms) or objects (MIT), $y$ the median scale-invariant error against the ground-truth albedo. Left: the drift ratio \eqref{eq:drift}; right: its contrast-matched form \eqref{eq:cm}. Top: OpenRooms (\TORn{} views, two lightings); bottom: MIT Intrinsics (20 objects, 10 lights). Lower plain drift accompanies higher error because the refinements lower the ratio by flattening; the contrast-matched form is positively correlated with the error. Coordinates are in the supplementary material.}
\label{fig:inv_err}
\end{figure*}

Figure~\ref{fig:inv_err} plots both forms of drift against the ground-truth error for the base models and the operating points of Section~\ref{sec:probes}. The plain ratio ranks them in the wrong order: across the \TORAtwoAllN{} points on OpenRooms the Spearman correlation between median drift and median scale-invariant error is \TORAtwoAllFive, across the \TMITAtwoAllN{} points on MIT it is \TMITAtwoAllFive; lower drift goes with \emph{higher} error. The contrast-matched form has the expected sign: \TORAtwoAllSix{} and \TMITAtwoAllSix. These points are not independent samples: eleven derive from one base model and two from another and share their bases' errors, so the correlations describe how refinements move relative to their base as much as how models differ. Restricted to the six base models ($n = 6$; descriptive correlations over the evaluated models, not estimates for a population of models) the correlations are \TORAtwoBaseFive{} and \TMITAtwoBaseFive{} for $\drift$ against \TORAtwoBaseSix{} and \TMITAtwoBaseSix{} for $\drift_z$. Resampling the scenes (\TORnScenesUsed{} scenes behind the \TORn{} views; objects on MIT) jointly for every method, \TAtwoNboot{} times with seed \TAtwoSeed, gives 95\% intervals of [\TORAtwoBaseFiveLo, \TORAtwoBaseFiveHi] for $\drift$ and [\TORAtwoBaseSixLo, \TORAtwoBaseSixHi] for $\drift_z$ on OpenRooms, and [\TMITAtwoBaseFiveLo, \TMITAtwoBaseFiveHi] and [\TMITAtwoBaseSixLo, \TMITAtwoBaseSixHi] on MIT; every correlation of this section keeps its sign in at least \TAtwoSignMin\% of the resamples except the $\drift$ correlation over the six base models on MIT (\TMITAtwoBaseFiveSign\%). (Scene-sampling uncertainty for the fixed model set, not generalization to other models.) The mechanism is visible in the contrast column of Table~\ref{tab:main} and in the next section: the operating points with the lowest drift ratios are the flattest albedos, and flat albedos are wrong.

\subsection{What a consistency objective does}
\label{sec:res_probes}
\begin{figure*}[t]
\centering
\includegraphics[width=0.70\linewidth]{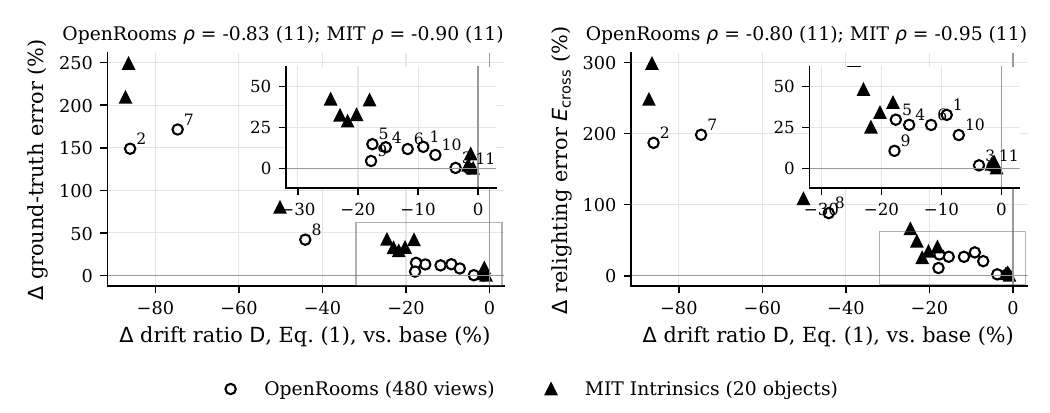}
\caption{Refinements of \ca{} v2 trained with a cross-illumination consistency objective (and two smoothing filters), relative to the base. Left: change of the median drift ratio against change of the mean ground-truth error; right: against change of the median relighting error $E_{\mathrm{cross}}$ of Section~\ref{sec:downstream}. Hollow circles: OpenRooms; filled triangles: MIT Intrinsics. Lowering the drift ratio raises both errors. Point keys (OpenRooms): \RefKey.}
\label{fig:refine}
\end{figure*}

Table~\ref{tab:ops} lists the operating points and Fig.~\ref{fig:refine} plots the refinements of \ca{} v2 against their base. On multilum the main adapter ($\lambda = 1$) lowers the drift ratio from \CAMLMed{} to \ADMLMed{} ($-$\ADMLRedMed{} of medians, $-$\ADMLRedMean{} of means) at no cost in full-IIW WHDR (\ADWhdr, \ADWhdrChange), the result a preliminary version of this work reported as an improvement. Contrast-matched, the same adapter changes drift by \ADMLCmChange{} [\ADMLCmChangeLo, \ADMLCmChangeHi] ($p\,\ADMLCmP$), and its albedo keeps \ADMLCon{} of the photograph's contrast against \CAMLCon{} for the base. On MIT its mean ground-truth error rises from \CAMITeSimLin{} to \ADMITeSimLin{} (\ADMITeSimChange, $p\,\ADMITeSimP$; the tables report medians), on OpenRooms by \TadORdSim\% (worse in \TadORfracWorse\% of views). The fine-tunes go further down the same path: at $\lambda = 1$ the drift ratio falls to \FtAMLMed{} while the albedo keeps \FtAMLCon{} of the photograph's contrast and the full-IIW WHDR rises by \FtAWhdrFullChange; at $\lambda = 3$ (\FtCMLMed, WHDR \FtCWhdrFullChange) the contrast-matched drift is \emph{worse} than the base's by \FtCMLCmChange. On NIID-Net the adapter lowers the ratio from \NIIDMLMed{} to \ADNMLMed{} ($-$\ADNMLRedMed) and the 300-image WHDR by \ADNWhdrChangeThree, while its MIT error rises by \ADNMITeSimChange{} and its contrast-matched drift on OpenRooms by \TadnORdCm\%.

Across the \TORAthreeN{} \ca{}-derived points (nine trained refinements, two filters) the change in the median drift ratio and the change in the \emph{mean} ground-truth error have a Spearman correlation of \TORAthree{} on OpenRooms and \TMITAthree{} on MIT: the more a refinement lowers the ratio, the more it raises the error (Fig.~\ref{fig:refine}, left). The refinements that leave the mean error within a few percent (the $\lambda = 3$ adapter, the bilateral filter) are the ones that barely change the drift ratio; the $\lambda = 10$ fine-tune is a milder trade-off (OpenRooms drift \TcaORDrift{} $\to$ \TftjORDrift, mean error \TftjORdSim\% on OpenRooms and \TftjMITdSim\% on MIT). The contrast-matched drift of the trained refinements is within a few percent of the base's or higher (Table~\ref{tab:ops}; OpenRooms \TCmRefORMin\% to \TCmRefORMax\%, multilum \TCmRefMLMin\% to \TCmRefMLMax\%, the largest decrease being the main adapter's at $p\,\ADMLCmP$; the two filters lower it by \TCmFiltORMin\% to \TCmFiltORMax\% on OpenRooms): the objective did not establish a contrast-matched stability improvement on multilum and raised $\drift_z$ on OpenRooms; it made the albedo flatter. The within-image score responded unevenly to the flattening, on the same outputs: on the \TiiwN{} IIW images of the WHDR subset the refinements compress contrast as much as on multilum (contrast ratio to the base albedo, median over images, within \TiiwMLConMaxDev{} of the multilum value for every refinement); the main adapter keeps \TiiwadCon{} at a WHDR change of \TiiwadWhdr\% and the $\lambda = 3$ fine-tune \TiiwftcCon{} at \TiiwftcWhdr\%, while the collapsed $\lambda = 0.3$ adapter (\TiiwlamaCon) and $\lambda = 1$ fine-tune (\TiiwftaCon) raise it by \TiiwlamaWhdr\% and \TiiwftaWhdr\%, and a Gaussian blur at a ratio of \TiiwblurCon{} by \TiiwblurWhdr\% (supplementary table): some substantial contrast reductions changed WHDR little, the collapsed outputs and the blur raised it, and which alterations the score registers in general would need controlled perturbations. Read alone, the drift ratio and a WHDR that changed by \TiiwadWhdr\% on this subset (\ADWhdrChange{} on the full set) would have reported the main adapter as an improvement.

\begin{table*}[t]
\centering
\caption{Refinement operating points with their bases (the other base models are in Table~\ref{tab:main}); all probes trained on multilum, Section~\ref{sec:probes}, plus two filters; refinements trained on other data belong to a separate study and are excluded. Multilum: drift \eqref{eq:drift}, contrast-matched drift \eqref{eq:cm} and albedo contrast relative to the photograph (medians over 30 scenes). OpenRooms (\TORn{} views): drift, $\drift_z$, median si-MSE, illumination-dependent fraction $f_{\mathrm{dep}}$, and the change of the median relighting error $E_{\mathrm{cross}}$ relative to the base. MIT (20 objects, 10 lights): median si-MSE, $f_{\mathrm{dep}}$, and the change of $E_{\mathrm{cross}}$. Trained refinements use the cross-illumination consistency objective of Section~\ref{sec:probes}; blur and bilateral are filters applied to the base albedo.}
\label{tab:ops}
\footnotesize\renewcommand{\arraystretch}{0.93}
\setlength{\tabcolsep}{4pt}
\begin{tabular}{l ccc cccc c ccc}
\toprule
 & \multicolumn{3}{c}{multilum} & \multicolumn{5}{c}{OpenRooms} & \multicolumn{3}{c}{MIT Intrinsics} \\
\cmidrule(lr){2-4} \cmidrule(lr){5-9} \cmidrule(lr){10-12}
Operating point & Drift & $\drift_z$ & contrast & Drift & $\drift_z$ & si-MSE & $f_{\mathrm{dep}}$ & $\Delta E_{\mathrm{cross}}$ & si-MSE & $f_{\mathrm{dep}}$ & $\Delta E_{\mathrm{cross}}$ \\
\midrule
\ca{} v2 (base) & \CAMLMed & \CAMLCm & \CAMLCon & \TcaORDrift & \TcaORCm & \TcaORSim & \TcaORFdep & --- & \TcaMITSim & \TcaMITFdep & --- \\
\quad + adapter $\lambda{=}1$ & \ADMLMed & \ADMLCm & \ADMLCon & \TadORDrift & \TadORCm & \TadORSim & \TadORFdep & \TadORdEcross\% & \TadMITSim & \TadMITFdep & \TadMITdEcross\% \\
\quad + adapter $\lambda{=}1$, retrained 10k & \ADtenkMLMed & \ADtenkMLCm & \ADtenkMLCon & \TadtenkORDrift & \TadtenkORCm & \TadtenkORSim & \TadtenkORFdep & \TadtenkORdEcross\% & \TadtenkMITSim & \TadtenkMITFdep & \TadtenkMITdEcross\% \\
\quad + adapter $\lambda{=}1$, retrained 20k & \ADrerunMLMed & \ADrerunMLCm & \ADrerunMLCon & \TadrerunORDrift & \TadrerunORCm & \TadrerunORSim & \TadrerunORFdep & \TadrerunORdEcross\% & \TadrerunMITSim & \TadrerunMITFdep & \TadrerunMITdEcross\% \\
\quad + adapter $\lambda{=}0.3$ & \LamAMLMed & \LamAMLCm & \LamAMLCon & \TlamaORDrift & \TlamaORCm & \TlamaORSim & \TlamaORFdep & \TlamaORdEcross\% & \TlamaMITSim & \TlamaMITFdep & \TlamaMITdEcross\% \\
\quad + adapter $\lambda{=}3$ & \LamCMLMed & \LamCMLCm & \LamCMLCon & \TlamcORDrift & \TlamcORCm & \TlamcORSim & \TlamcORFdep & \TlamcORdEcross\% & \TlamcMITSim & \TlamcMITFdep & \TlamcMITdEcross\% \\
\quad + adapter, albedo-only input & \NoImgMLMed & \NoImgMLCm & \NoImgMLCon & \TnoimgORDrift & \TnoimgORCm & \TnoimgORSim & \TnoimgORFdep & \TnoimgORdEcross\% & \TnoimgMITSim & \TnoimgMITFdep & \TnoimgMITdEcross\% \\
\quad fine-tuned, $\lambda{=}1$ & \FtAMLMed & \FtAMLCm & \FtAMLCon & \TftaORDrift & \TftaORCm & \TftaORSim & \TftaORFdep & \TftaORdEcross\% & \TftaMITSim & \TftaMITFdep & \TftaMITdEcross\% \\
\quad fine-tuned, $\lambda{=}3$ & \FtCMLMed & \FtCMLCm & \FtCMLCon & \TftcORDrift & \TftcORCm & \TftcORSim & \TftcORFdep & \TftcORdEcross\% & \TftcMITSim & \TftcMITFdep & \TftcMITdEcross\% \\
\quad fine-tuned, $\lambda{=}10$ & \FtJMLMed & \FtJMLCm & \FtJMLCon & \TftjORDrift & \TftjORCm & \TftjORSim & \TftjORFdep & \TftjORdEcross\% & \TftjMITSim & \TftjMITFdep & \TftjMITdEcross\% \\
\quad + Gaussian blur $\sigma{=}4$ & \BlurMLMed & \BlurMLCm & \BlurMLCon & \TblurORDrift & \TblurORCm & \TblurORSim & \TblurORFdep & \TblurORdEcross\% & \TblurMITSim & \TblurMITFdep & \TblurMITdEcross\% \\
\quad + bilateral filter & \BilatMLMed & \BilatMLCm & \BilatMLCon & \TbilatORDrift & \TbilatORCm & \TbilatORSim & \TbilatORFdep & \TbilatORdEcross\% & \TbilatMITSim & \TbilatMITFdep & \TbilatMITdEcross\% \\
\midrule
NIID-Net (base) & \NIIDMLMed & \NIIDMLCm & \NIIDMLCon & \TniidORDrift & \TniidORCm & \TniidORSim & \TniidORFdep & --- & \TniidMITSim & \TniidMITFdep & --- \\
\quad + adapter $\lambda{=}1$ & \ADNMLMed & \ADNMLCm & \ADNMLCon & \TadnORDrift & \TadnORCm & \TadnORSim & \TadnORFdep & \TadnORdEcross\% & \TadnMITSim & \TadnMITFdep & \TadnMITdEcross\% \\
\quad + adapter $\lambda{=}1$, 10k & \ADNkMLMed & \ADNkMLCm & \ADNkMLCon & \TadnkORDrift & \TadnkORCm & \TadnkORSim & \TadnkORFdep & \TadnkORdEcross\% & --- & --- & --- \\
\bottomrule
\end{tabular}
\end{table*}

\subsection{The illumination-dependent fraction of the error}
\label{sec:res_fdep}
\begin{figure*}[t]
\centering
\includegraphics[width=0.68\linewidth]{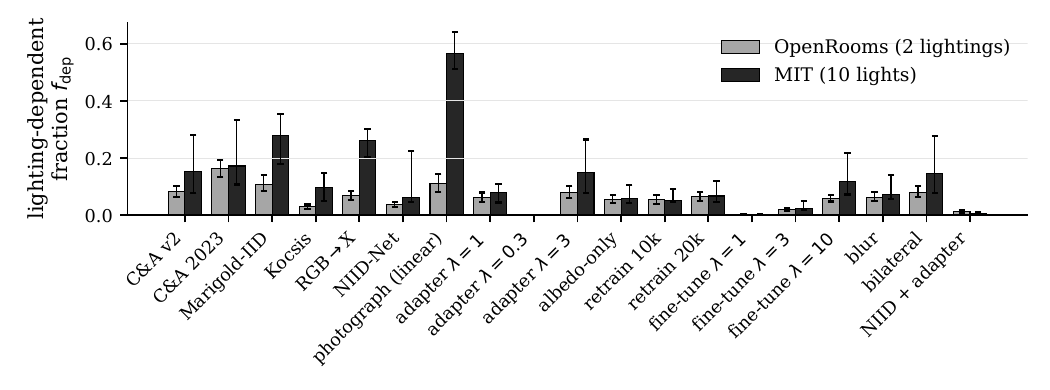}
\caption{Illumination-dependent fraction $f_{\mathrm{dep}}$ of the ground-truth albedo error, \eqref{eq:fdep}, per method (medians, 95\% intervals from a bootstrap over scenes with all their views, or over objects); ``photograph (linear)'' is the linear photograph taken as the albedo. Grey: OpenRooms, two lightings per view; black: MIT Intrinsics, ten lights per object.}
\label{fig:fdep}
\end{figure*}

Figure~\ref{fig:fdep} reports $f_{\mathrm{dep}}$ of \eqref{eq:fdep}. On the OpenRooms pairs the illumination-dependent part is \TORFdepBaseMin--\TORFdepBaseMax\% of the ground-truth error of the base models (\ca{} v2 \TcaORFdep{} [\TcaORFdepLo, \TcaORFdepHi], Marigold-IID \TmgORFdep, Kocsis et al.\ \TkoORFdep), and \TidlORFdep{} for the linear photograph itself (\TidORFdep{} for the display-encoded photograph): the two lightings of this dataset share most of their shading structure, so even the photograph's deviation from albedo is mostly common to both. On MIT, whose ten directional lights move shadows and shading across the object, it is \TMITFdepBaseMin--\TMITFdepBaseMax\% for the base models (Marigold-IID \TmgMITFdep, RGB$\to$X \TrgbxMITFdep, \ca{} v2 \TcaMITFdep{} [\TcaMITFdepLo, \TcaMITFdepHi]) and \TidlMITFdep{} for the linear photograph (\TidMITFdep{} encoded). The most collapsed refinements, the $\lambda = 0.3$ adapter and the $\lambda = 1$ fine-tune, have $f_{\mathrm{dep}} \leq 0.002$: they removed the illumination-dependent component and replaced it with a larger shared bias; for the milder refinements the fraction falls less (main adapter \TadORFdep{} on OpenRooms, \TadMITFdep{} on MIT) while the shared error still grows (Table~\ref{tab:ops}). Drift therefore tracks, without ground truth, a part of the error whose size differs between datasets, from a few percent for the two indoor configurations to a quarter for the ten directional lights; for every learned operating point the larger part is a residual shared across the evaluated lightings (the photograph on MIT is the exception, \TidlMITFdep). The datasets differ in more than lighting diversity (content, materials, $L$ in \eqref{eq:identity}), so attributing the difference to the lightings is a plausible reading, not an identified cause.

\subsection{Downstream: re-using an albedo across lightings}
\label{sec:res_down}
On OpenRooms the excess relighting error of Section~\ref{sec:downstream} is small for every learned operating point: the median of $\Delta E / E_{\mathrm{within}}$ lies between \TORBoneMin{} and \TORBoneMax{} over the \TORBoneN{} points (base \ca{} v2 \TcaORBoneF{} [\TcaORBoneLoF, \TcaORBoneHiF]); \TORBoneCIexcl{} of the intervals exclude zero, but no endpoint exceeds \TORBoneCIabsMax{} in magnitude, against a pre-registered expectation of 0.15--0.60, and the photograph used as albedo gives \TidORBone. An albedo estimated under the other lighting has a small median excess relighting error under this protocol, consistent with the small illumination-dependent fraction, although the relighting score weights and aligns the residual differently (Section~\ref{sec:downstream}). On MIT, where the fraction is larger, the cost appears: $\Delta E / E_{\mathrm{within}}$ is \TMITBoneBaseMin--\TMITBoneBaseMax\% for the base models (RGB$\to$X \TrgbxMITBone{} [\TrgbxMITBoneLo, \TrgbxMITBoneHi], \ca{} v2 \TcaMITBone{} [\TcaMITBoneLo, \TcaMITBoneHi], Marigold-IID \TmgMITBone, NIID-Net \TniidMITBone; photograph as albedo \TidMITBone). The refinements that lower the drift ratio are worse at the task on both datasets: $E_{\mathrm{cross}}$ exceeds the base's by \TORdEcrossMin\% to \TORdEcrossMax\% on OpenRooms (worse in \TORfracEcrossWorseMin--\TORfracEcrossWorseMax\% of views) and \TMITdEcrossMin\% to \TMITdEcrossMax\% on MIT, with the $\lambda = 1$ fine-tune and the $\lambda = 0.3$ adapter at the top (Table~\ref{tab:ops}, Fig.~\ref{fig:refine} right): a flat albedo times the correct shading does not reproduce the photograph.

\subsection{Protocol sensitivities}
\label{sec:res_protocol}
\textbf{Encoding.} Under the mixed convention (linear output against encoded photograph) the base models' multilum drift ratios would be \TMixOverMatchedMin--\TMixOverMatchedMax{} times larger and a no-op would score \IDLINMLMed{} (Section~\ref{sec:method}); matched encoding and the identity control are not optional. \textbf{Input linearization.} Running the model of \cite{kocsis24iid} with an exact sRGB linearization instead of its authors' exponent $2.2$ changes its multilum drift by \KOgMLDiffMed{} (within the interval). \textbf{Exposure policy.} The OpenRooms renders are linear radiance and must be exposed before 8-bit encoding; the protocol fixes one exposure per view (a fixed camera). Re-running every model with each frame exposed separately (an auto-exposure camera), compared on the same pixels (per-method table in the supplement), changes the base models' ground-truth error by \TXSimBaseMin\% to \TXSimBaseMax\% and that of the photograph by \TXSimIdentity\%, but lowers the base models' drift ratio by \TXDriftBaseMin\% to \TXDriftBaseMax\% and the refinements' by \TXDriftRefMin\% to \TXDriftRefMax\% (\ca{} v2 on the common pixels: \TcaXDriftShared{} $\to$ \TcaXDriftPf). Every correlation of Section~\ref{sec:res_corr} keeps its sign under both policies (\TPFAtwoAllFive{} and \TPFAtwoAllSix{} across operating points per frame), but the measured benefit of an adapter depends on the exposure given to the photographs: a drift protocol must state its exposure rule.

\subsection{Pre-registered hypotheses that failed}
\label{sec:failed}
Three hypotheses were written down with a failure condition before the ground-truth experiments ran; all three met it. (i) We expected the contrast-matched drift $\drift_z$ of a view to predict, within each base model, how much the ground-truth error changes between its two lightings, $|e_a - e_b| / \bar{e}$, with a Spearman correlation of $0.35$--$0.70$ over views; failure: below $0.30$ with an upper confidence limit below $0.50$ for every base model. On OpenRooms the base models give \TORAoneBaseMin--\TORAoneBaseMax{} (upper limits at most \TORAoneBaseCiHiMax); on MIT, where the target is the coefficient of variation of the error over the ten lights, $\mathrm{std}_\ell(e_\ell)/\mathrm{mean}_\ell(e_\ell)$ over the 20 objects, \TMITAoneBaseMin--\TMITAoneBaseMax{} (at most \TMITAoneBaseCiHiMax). The refinements, not covered by the rule, span \TORAoneMin--\TORAoneMax{} and \TMITAoneRefMin--\TMITAoneRefMax; on MIT \TMITAoneRefArgmax{} reaches \TMITAoneRefMax{} [\TMITAoneRefMaxLo, \TMITAoneRefMaxHi], above the rule's limit (supplementary table). The hypothesis was badly posed: for two lightings $\varepsilon_{a,b} = \bar\varepsilon \pm \delta$, so $e_a - e_b = 4\langle\bar\varepsilon, \delta\rangle$ and $|e_a - e_b| \leq 4\|\bar\varepsilon\|\|\delta\|$: the sensitivity depends on the interaction of the shared and the varying residual, not on $\|\delta\|$ alone; two estimates can differ a great deal while being equally wrong. (ii) We expected the excess relighting error $\Delta E$ to be 15--60\% of $E_{\mathrm{within}}$ on OpenRooms; the operating-point medians are at most \TORBoneAbsMax{} in magnitude and no interval endpoint exceeds \TORBoneCIabsMax{} (Section~\ref{sec:res_down}). On MIT, for which we had written the opposite expectation, it is \TMITBoneBaseMin--\TMITBoneBaseMax\%; the two outcomes are consistent once the illumination-dependent fraction is known, but neither was predicted. (iii) We expected the contrast-matched drift of a view to predict its excess relighting error $\Delta E/E_{\mathrm{within}}$ with a Spearman correlation of $0.4$--$0.75$; across the base models it is \TORBtwoMin--\TORBtwoMax{} on OpenRooms, where the excess is itself near zero, and \TMITBtwoMin--\TMITBtwoMax{} over the \TMITBtwoN{} MIT objects, the wrong sign. Nor does the between-operating-point version hold: across the \TMITBtwoMethodsN{} MIT operating points the Spearman correlation between median $\drift_z$ and median $\Delta E/E_{\mathrm{within}}$ is \TMITBtwoMethods{} (\TMITBtwoMethodsBase{} over the six base models), because a flattened albedo has a small \emph{excess} cost (it is nearly the same under every light) but a large \emph{total} error. Flattening raises the total error (Section~\ref{sec:res_down}); the excess cost of re-use is a different quantity, which $\drift_z$ did not predict. We report them because an evaluation paper that reported only the hypotheses that survived would commit the error it describes.

\subsection{Qualitative results}
\begin{figure*}[!t]
\centering
\includegraphics[width=0.76\linewidth]{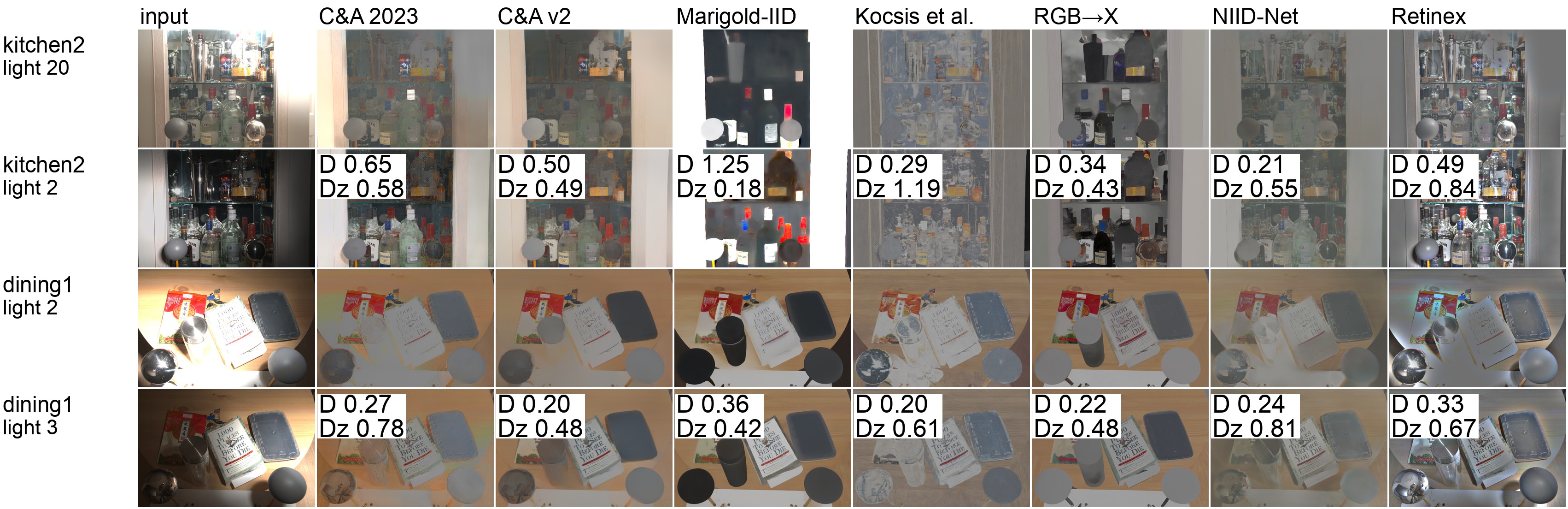}
\caption{Two multilum scenes under their two most different lights (rows; the pair of the 25 whose median-normalized photographs differ most, labelled with the dataset's light numbers). Columns: input and the methods of Table~\ref{tab:main}; albedos median-normalized and display-encoded. The light-b panel carries the scene's $\drift$ and $\drift_z$ over all 25 lightings. Four further scenes, with the refinements of Section~\ref{sec:probes}, are in the supplementary material.}
\label{fig:qual}
\end{figure*}

\begin{figure*}[!t]
\centering
\includegraphics[width=0.74\linewidth]{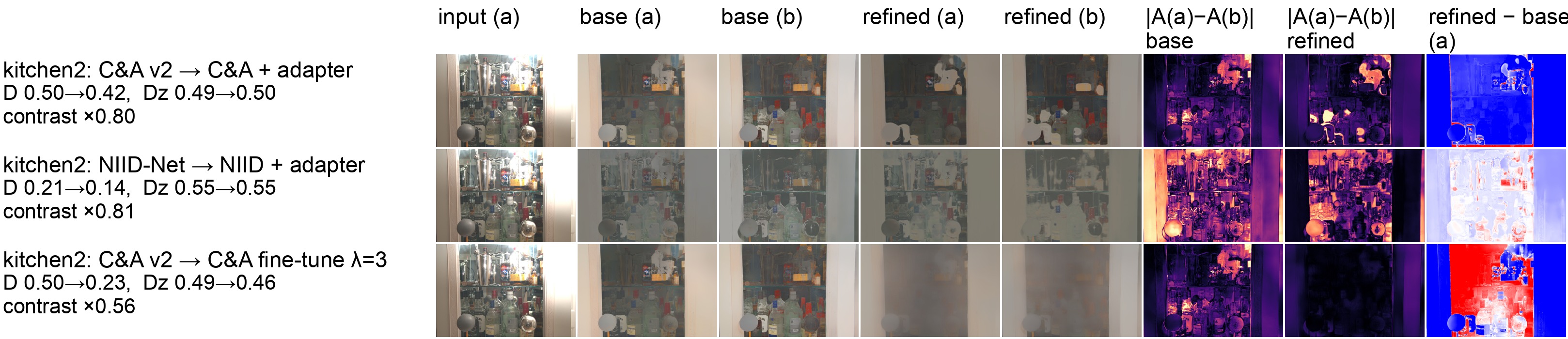}
\caption{What the consistency objective changes. For one scene (two more in the supplementary material): base and refined albedo under two lights, the per-pixel difference $|A_a - A_b|$ of each as a heat map with one scale per row (set by the base), and the signed residual (refined $-$ base) under light a on a fixed blue--white--red scale of $\pm 0.25$; heat maps and residual are channel means of the absolute and of the signed difference in median-normalized display-encoded units (opposing colour changes can cancel in the signed map), thumbnails the same units scaled to a median of $0.45$; row labels give the scene-level $\drift$, $\drift_z$ (over all 25 lightings) and contrast ratio. The refined albedo differs less between the lights because dark regions were brightened and bright regions darkened.}
\label{fig:qual_adapter}
\end{figure*}

Figure~\ref{fig:qual} shows two multilum scenes under their two most different lights for every method of Table~\ref{tab:main}, annotated with the scene's drift and contrast-matched drift over all 25 lightings (four more scenes, with the refinements, are in the supplementary material); the drift appears as a change of albedo brightness and colour between the two rows of a scene, concentrated at shadowed, specular and glass regions. Figure~\ref{fig:qual_adapter} shows what the adapter and the fine-tune change, as difference maps between the two lightings and as residuals against the base; the flattening appears as a loss of texture that the drift ratio rewards.

\section{Discussion}
\label{sec:discussion}
\textbf{What to report.} An evaluation with photographs of one scene under several lights can report three numbers without ground truth: the drift ratio, its contrast-matched form and the albedo's contrast relative to the photograph. Together they expose a flattened albedo (low $\drift$, $\drift_z$ unchanged, contrast reduced) behind a low ratio; they do not certify a stable one, since a low-contrast pattern repeated under every light passes both forms, and a correct albedo need not match the photograph's contrast, so contrast is reported, not scored. Where ground truth exists, $f_{\mathrm{dep}}$ sizes the varying component within the evaluated lighting set; fidelity scores assess overall recovery and remain necessary, since shared errors are invisible to cross-lighting differences.

\textbf{What not to optimize.} The result of Section~\ref{sec:res_probes} is an instance of a general hazard: a necessary condition, optimized alone, is satisfied through whatever degree of freedom the objective leaves open; here, contrast, to which WHDR and the drift ratio are both weakly sensitive. Consistency losses are a legitimate training signal as one of several terms of a network trained from scratch, as in \cite{li18bigtime}, which learns from time-lapses without ground-truth decompositions; applied to a trained model with only an anchor to the base, as in \eqref{eq:probe}, they trade the illumination-dependent error for a larger shared one. Lower drift alone should not be read as a better decomposition: $\drift$, $\drift_z$, contrast and fidelity should be reported separately with their trade-offs; a fidelity claim needs a fidelity measure, and less standardized cross-lighting variation is not better material recovery. A joint fall of $\drift$ and $\drift_z$ is conservative evidence of an invariance gain, neither necessary (a refinement that restores contrast can raise $\drift$ while lowering $\drift_z$) nor proof against collapse (Section~\ref{sec:cm}). Contrast itself has no preferred direction.

\textbf{Why the fraction is small, and when it is not.} On the indoor pairs $f_{\mathrm{dep}}$ is \TORFdepBaseMin--\TORFdepBaseMax\% for the published base models and \TidlORFdep{} for the photograph, whose two lightings share most of their shading structure; on lab objects under moving directional lights it is \TMITFdepBaseMin--\TMITFdepBaseMax\% for the published base models and \TidlMITFdep{} for the photograph, for which the illumination-dependent component is the majority. The downstream cost of re-use is large where the fraction is large and small in the OpenRooms summaries where it is small, without an established quantitative law between the two. For applications that mix captures under strongly different lights (relighting, material capture across a day, video under changing illumination) the leak measured here contributes to the error that matters, although which component dominates in practice is not established; under similar lights the shared residual dominates and only ground-truth error or human judgement sees it.

\section{Limitations}
(1) The drift ratio is a necessary condition; its contrast-matched form and a ground-truth anchor are needed to read it, and no method improving it is proposed. (2) Ground truth exists only on synthetic renders and 20 lab objects; real photographs support the drift measurement, not the decomposition. (3) The downstream measurement uses the ground-truth shading and a least-squares scale; it isolates the albedo's contribution to relighting but is not a user-facing application, and no user study was run. (4) The contrast-matched form standardizes one spatial statistic; a refinement that keeps global contrast while removing texture at one scale would pass it. (5) The image-formation model $I = R\odot S$ puts interreflections and specularities into $S$; the exact albedo scores $\drift = 0$ regardless (the control holds), and a decomposition that absorbs unmodelled transport into its albedo is charged for it; the diffuse-only mask rejects saturated and moving highlights, not all non-Lambertian effects, so the ratios measure the invariance failure of the \emph{estimated} layer, not a separation of physical causes. (6) The exposure study covers one synthetic dataset and two policies; real cameras apply tone curves and local adjustments that neither emulates. (7) The spectral component of the illumination change is not measured separately, and on real photographs the drift of an estimated albedo does not separate an algorithm's failure from the limits of the RGB factorization (Section~\ref{sec:method}); only on the renders is it exact. (8) All probes of the consistency objective share one dataset and one recipe; other recipes may behave differently, and the claim is only that WHDR alone failed to reject refinements whose fidelity had worsened.

\section{Conclusion}
\label{sec:conclusion}
Albedo drift measures the property that defines albedo and that the field's evaluation protocols do not test. Its numerator is a ground-truth-free surrogate of the illumination-dependent component of the albedo error, a modest part of the whole for the published models, \TORFdepBaseMin--\TMITFdepBaseMax\% across the two ground-truth datasets (smaller still for the collapsed refinements), and the part accessible without ground truth through cross-lighting differences. Optimizing it alone is harmful: across \TORAtwoAllN{} operating points lower plain drift accompanies higher ground-truth error, because a consistency objective applied to a trained model lowers the ratio by flattening, and the relighting task that motivates invariance gets worse. Reporting drift with its contrast-matched form, the albedo's contrast and, where possible, ground-truth error gives the field a measurement of invariance that exposes rather than hides contrast loss; by it, none of the evaluated models is invariant, and independent fidelity evidence remains necessary. Code, pre-registration records and result files will be released.

\section*{Acknowledgment}
AI-assisted tools were used to write and debug the experiment code and to edit the English; all design decisions, pre-registrations, interpretations and claims are the author's.

\bibliographystyle{IEEEtran}
\bibliography{refs}

\clearpage
\setcounter{dbltopnumber}{3}\renewcommand{\dbltopfraction}{0.95}\renewcommand{\dblfloatpagefraction}{0.85}
\twocolumn[{\centering {\LARGE\bfseries Supplementary Material}\\[0.8ex]
{\large for ``Albedo Drift: Measuring the Illumination-Dependent Error of Intrinsic Image Decomposition''}\par\vspace{2em}}]

\section{Pre-registration and outcomes}
Every ground-truth experiment was written down with its quantity, expected range and failure condition before the result files were read. Table~\ref{tab:prereg} lists them with the outcome.

\begin{table*}[t]
\centering
\caption{Pre-registered hypotheses of the ground-truth experiments and their outcomes.}
\label{tab:prereg}
\small
\begin{tabular}{p{0.24\linewidth} p{0.24\linewidth} p{0.40\linewidth}}
\toprule
Hypothesis & Expected range / failure condition & Outcome \\
\midrule
H-A1: view-level Spearman, within each base model, between contrast-matched drift $\drift_z$ and the lighting sensitivity of the ground-truth error ($|e_a-e_b|/\bar e$ over the two OpenRooms lightings; on MIT, written after the OpenRooms result, $\mathrm{std}_\ell(e_\ell)/\mathrm{mean}_\ell(e_\ell)$ over ten lights and 20 objects) & $0.35$--$0.70$ per base model; failure if every base model $<0.30$ with upper CI limit $<0.50$; ``too strong'' if $>0.90$ & \textbf{Failed} for the base models on OpenRooms (\TORAoneBaseMin--\TORAoneBaseMax, upper limits $\leq$ \TORAoneBaseCiHiMax) and on MIT (\TMITAoneBaseMin--\TMITAoneBaseMax, $\leq$ \TMITAoneBaseCiHiMax). Refinements, not covered by the rule: \TORAoneMin--\TORAoneMax{} on OpenRooms; \TMITAoneRefMin--\TMITAoneRefMax{} on MIT, where \TMITAoneRefArgmax{} reaches \TMITAoneRefMax{} [\TMITAoneRefMaxLo, \TMITAoneRefMaxHi]. Diagnosis: $e_a - e_b = 4\langle\bar\varepsilon, \delta\rangle$, so the sensitivity depends on the interaction of the shared and the varying residual, not on $\|\delta\|$ alone. \\
H-A2: method-level Spearman between median $\drift_z$ and median si-MSE over the base models & $>0.60$; the same with $\drift$ expected lower & Held: \TORAtwoBaseSix{} (OpenRooms), \TMITAtwoBaseSix{} (MIT); with $\drift$: \TORAtwoBaseFive, \TMITAtwoBaseFive. \\
H-A3: refinements of \ca{}: sign of (drift change, error change) & negative association (MIT pilot in the preliminary version gave $-0.79$) & Held: \TORAthree{} (OpenRooms), \TMITAthree{} (MIT), \TORAthreeN{} refinements. \\
H-B1: excess relighting error $\Delta E / E_{\mathrm{within}}$ of the base models & OpenRooms: $0.15$--$0.60$; failure if every learned model $<0.10$. MIT (written after the OpenRooms result): $|\Delta E/E_{\mathrm{within}}| < 0.05$ & \textbf{Failed} on OpenRooms in the direction of a much smaller cost (medians \TORBoneMin--\TORBoneMax; \TORBoneCIexcl{} of \TORBoneN{} intervals exclude zero, none beyond \TORBoneCIabsMax); the MIT expectation was contradicted in the direction of a cost (\TMITBoneBaseMin--\TMITBoneBaseMax\% for the base models). \\
H-B2: view-level Spearman between $\drift_z$ and $\Delta E/E_{\mathrm{within}}$, and agreement of the method ranking with the $\drift_z$ ranking & $0.4$--$0.75$ per method & \textbf{Failed}: base models \TORBtwoMin--\TORBtwoMax{} on OpenRooms (where $\Delta E/E_{\mathrm{within}}$ is near zero for every method) and \TMITBtwoMin--\TMITBtwoMax{} on MIT; the ranking part is moot on OpenRooms and was not met on MIT. \\
H-B3: refinements' $E_{\mathrm{cross}}$ relative to the base & not lower & Held: \TORdEcrossMin\% to \TORdEcrossMax\% (OpenRooms), \TMITdEcrossMin\% to \TMITdEcrossMax\% (MIT). \\
$f_{\mathrm{dep}}$ of the base models & OpenRooms $0.15$--$0.45$ (written before the run; failure if all $<0.05$); MIT $0.10$--$0.40$ & OpenRooms \TORFdepBaseMin--\TORFdepBaseMax\% (the expectation was too high; no failure); MIT \TMITFdepBaseMin--\TMITFdepBaseMax\%; flattened refinements $<0.10$ and Marigold-IID $>$ \ca{} held on both. \\
Exposure policy (per-frame vs.\ shared) & base-model drift change $<15\%$ and si-MSE change $<10\%$ $\Rightarrow$ ``insensitive''; any model $>30\%$ $\Rightarrow$ report the dependence & Not insensitive: drift changes \TXDriftBaseMin\% to \TXDriftBaseMax\% (base), \TXDriftRefMin\% to \TXDriftRefMax\% (refinements) on the common pixels; si-MSE changes \TXSimBaseMin\% to \TXSimBaseMax\%. The dependence is reported (Section~V-F of the article and Table~\ref{tab:exposure} of this supplement). \\
\bottomrule
\end{tabular}
\end{table*}

\section{Per-method statistics on OpenRooms}
Table~\ref{tab:supp_or} lists, for every operating point, the view-level quantities behind Figs.~1--3 and Sections~V-B to V-E of the article: medians over the \TORn{} views of the drift ratio, the contrast-matched drift, the albedo contrast relative to the photograph, the scale-invariant error against the ground truth, the lighting sensitivity of that error, the illumination-dependent fraction with its 95\% scene-bootstrap interval, the view-level Spearman correlation of H-A1 with its interval, the excess relighting error with its interval, and the Spearman correlation between the drift of mismatched views (albedo of one view against albedo of another view of the same scene) and the lighting sensitivity, the tested diagnostic of Section~III-G ($|\rho| < 0.15$).

\begin{table*}[t]
\centering
\caption{OpenRooms, \TORn{} views. Medians; brackets are 95\% scene-bootstrap intervals.}
\label{tab:supp_or}
\footnotesize
\setlength{\tabcolsep}{2.6pt}
\begin{tabular}{l cccc c c c c c}
\toprule
Operating point & Drift & $\drift_z$ & contrast & si-MSE & sens. & $f_{\mathrm{dep}}$ [CI] & $\rho(\drift_z,\mathrm{sens})$ [CI] & $\Delta E/E_{\mathrm{within}}$ [CI] & $\rho$(mismatched) \\
\midrule
\ca{} v2 & \TcaORDrift & \TcaORCm & \TcaORCon & \TcaORSim & \TcaORSens & \TcaORFdep{} [\TcaORFdepLo, \TcaORFdepHi] & \TcaORAone{} [\TcaORAoneLo, \TcaORAoneHi] & \TcaORBone{} [\TcaORBoneLo, \TcaORBoneHi] & \TcaORRp \\
\ca{} 2023 & \TcaoORDrift & \TcaoORCm & \TcaoORCon & \TcaoORSim & \TcaoORSens & \TcaoORFdep{} [\TcaoORFdepLo, \TcaoORFdepHi] & \TcaoORAone{} [\TcaoORAoneLo, \TcaoORAoneHi] & \TcaoORBone{} [\TcaoORBoneLo, \TcaoORBoneHi] & \TcaoORRp \\
NIID-Net & \TniidORDrift & \TniidORCm & \TniidORCon & \TniidORSim & \TniidORSens & \TniidORFdep{} [\TniidORFdepLo, \TniidORFdepHi] & \TniidORAone{} [\TniidORAoneLo, \TniidORAoneHi] & \TniidORBone{} [\TniidORBoneLo, \TniidORBoneHi] & \TniidORRp \\
Kocsis et al. & \TkoORDrift & \TkoORCm & \TkoORCon & \TkoORSim & \TkoORSens & \TkoORFdep{} [\TkoORFdepLo, \TkoORFdepHi] & \TkoORAone{} [\TkoORAoneLo, \TkoORAoneHi] & \TkoORBone{} [\TkoORBoneLo, \TkoORBoneHi] & \TkoORRp \\
RGB$\to$X & \TrgbxORDrift & \TrgbxORCm & \TrgbxORCon & \TrgbxORSim & \TrgbxORSens & \TrgbxORFdep{} [\TrgbxORFdepLo, \TrgbxORFdepHi] & \TrgbxORAone{} [\TrgbxORAoneLo, \TrgbxORAoneHi] & \TrgbxORBone{} [\TrgbxORBoneLo, \TrgbxORBoneHi] & \TrgbxORRp \\
Marigold-IID & \TmgORDrift & \TmgORCm & \TmgORCon & \TmgORSim & \TmgORSens & \TmgORFdep{} [\TmgORFdepLo, \TmgORFdepHi] & \TmgORAone{} [\TmgORAoneLo, \TmgORAoneHi] & \TmgORBone{} [\TmgORBoneLo, \TmgORBoneHi] & \TmgORRp \\
Retinex (log-domain) & \TrxORDrift & \TrxORCm & \TrxORCon & \TrxORSim & \TrxORSens & \TrxORFdep{} [\TrxORFdepLo, \TrxORFdepHi] & \TrxORAone{} [\TrxORAoneLo, \TrxORAoneHi] & \TrxORBone{} [\TrxORBoneLo, \TrxORBoneHi] & --- \\
Identity & 1.000 & 1.000 & 1.00 & \TidORSim & \TidORSens & \TidORFdep{} [\TidORFdepLo, \TidORFdepHi] & 0.00 & \TidORBone{} [\TidORBoneLo, \TidORBoneHi] & --- \\
\midrule
\ca{} + adapter $\lambda{=}1$ & \TadORDrift & \TadORCm & \TadORCon & \TadORSim & \TadORSens & \TadORFdep{} [\TadORFdepLo, \TadORFdepHi] & \TadORAone{} [\TadORAoneLo, \TadORAoneHi] & \TadORBone{} [\TadORBoneLo, \TadORBoneHi] & \TadORRp \\
\ca{} + adapter, retrained 10k & \TadtenkORDrift & \TadtenkORCm & \TadtenkORCon & \TadtenkORSim & \TadtenkORSens & \TadtenkORFdep{} [\TadtenkORFdepLo, \TadtenkORFdepHi] & \TadtenkORAone{} [\TadtenkORAoneLo, \TadtenkORAoneHi] & \TadtenkORBone{} [\TadtenkORBoneLo, \TadtenkORBoneHi] & \TadtenkORRp \\
\ca{} + adapter, retrained 20k & \TadrerunORDrift & \TadrerunORCm & \TadrerunORCon & \TadrerunORSim & \TadrerunORSens & \TadrerunORFdep{} [\TadrerunORFdepLo, \TadrerunORFdepHi] & \TadrerunORAone{} [\TadrerunORAoneLo, \TadrerunORAoneHi] & \TadrerunORBone{} [\TadrerunORBoneLo, \TadrerunORBoneHi] & \TadrerunORRp \\
\ca{} + adapter $\lambda{=}0.3$ & \TlamaORDrift & \TlamaORCm & \TlamaORCon & \TlamaORSim & \TlamaORSens & \TlamaORFdep{} [\TlamaORFdepLo, \TlamaORFdepHi] & \TlamaORAone{} [\TlamaORAoneLo, \TlamaORAoneHi] & \TlamaORBone{} [\TlamaORBoneLo, \TlamaORBoneHi] & \TlamaORRp \\
\ca{} + adapter $\lambda{=}3$ & \TlamcORDrift & \TlamcORCm & \TlamcORCon & \TlamcORSim & \TlamcORSens & \TlamcORFdep{} [\TlamcORFdepLo, \TlamcORFdepHi] & \TlamcORAone{} [\TlamcORAoneLo, \TlamcORAoneHi] & \TlamcORBone{} [\TlamcORBoneLo, \TlamcORBoneHi] & \TlamcORRp \\
\ca{} + adapter, albedo-only & \TnoimgORDrift & \TnoimgORCm & \TnoimgORCon & \TnoimgORSim & \TnoimgORSens & \TnoimgORFdep{} [\TnoimgORFdepLo, \TnoimgORFdepHi] & \TnoimgORAone{} [\TnoimgORAoneLo, \TnoimgORAoneHi] & \TnoimgORBone{} [\TnoimgORBoneLo, \TnoimgORBoneHi] & \TnoimgORRp \\
\ca{} fine-tuned $\lambda{=}1$ & \TftaORDrift & \TftaORCm & \TftaORCon & \TftaORSim & \TftaORSens & \TftaORFdep{} [\TftaORFdepLo, \TftaORFdepHi] & \TftaORAone{} [\TftaORAoneLo, \TftaORAoneHi] & \TftaORBone{} [\TftaORBoneLo, \TftaORBoneHi] & \TftaORRp \\
\ca{} fine-tuned $\lambda{=}3$ & \TftcORDrift & \TftcORCm & \TftcORCon & \TftcORSim & \TftcORSens & \TftcORFdep{} [\TftcORFdepLo, \TftcORFdepHi] & \TftcORAone{} [\TftcORAoneLo, \TftcORAoneHi] & \TftcORBone{} [\TftcORBoneLo, \TftcORBoneHi] & \TftcORRp \\
\ca{} fine-tuned $\lambda{=}10$ & \TftjORDrift & \TftjORCm & \TftjORCon & \TftjORSim & \TftjORSens & \TftjORFdep{} [\TftjORFdepLo, \TftjORFdepHi] & \TftjORAone{} [\TftjORAoneLo, \TftjORAoneHi] & \TftjORBone{} [\TftjORBoneLo, \TftjORBoneHi] & \TftjORRp \\
\ca{} + Gaussian blur & \TblurORDrift & \TblurORCm & \TblurORCon & \TblurORSim & \TblurORSens & \TblurORFdep{} [\TblurORFdepLo, \TblurORFdepHi] & \TblurORAone{} [\TblurORAoneLo, \TblurORAoneHi] & \TblurORBone{} [\TblurORBoneLo, \TblurORBoneHi] & --- \\
\ca{} + bilateral & \TbilatORDrift & \TbilatORCm & \TbilatORCon & \TbilatORSim & \TbilatORSens & \TbilatORFdep{} [\TbilatORFdepLo, \TbilatORFdepHi] & \TbilatORAone{} [\TbilatORAoneLo, \TbilatORAoneHi] & \TbilatORBone{} [\TbilatORBoneLo, \TbilatORBoneHi] & --- \\
NIID-Net + adapter & \TadnORDrift & \TadnORCm & \TadnORCon & \TadnORSim & \TadnORSens & \TadnORFdep{} [\TadnORFdepLo, \TadnORFdepHi] & \TadnORAone{} [\TadnORAoneLo, \TadnORAoneHi] & \TadnORBone{} [\TadnORBoneLo, \TadnORBoneHi] & \TadnORRp \\
NIID-Net + adapter, 10k & \TadnkORDrift & \TadnkORCm & \TadnkORCon & \TadnkORSim & \TadnkORSens & \TadnkORFdep{} [\TadnkORFdepLo, \TadnkORFdepHi] & \TadnkORAone{} [\TadnkORAoneLo, \TadnkORAoneHi] & \TadnkORBone{} [\TadnkORBoneLo, \TadnkORBoneHi] & \TadnkORRp \\
\bottomrule
\end{tabular}
\end{table*}

\section{Per-method statistics on MIT Intrinsics}
Table~\ref{tab:supp_mit} gives the corresponding quantities on the 20 objects under 10 lights (90 ordered light pairs per object for the relighting measure).

\begin{table*}[t]
\centering
\caption{MIT Intrinsics, 20 objects $\times$ 10 lights. Medians over objects; brackets are 95\% bootstrap intervals over objects.}
\label{tab:supp_mit}
\footnotesize
\setlength{\tabcolsep}{4pt}
\begin{tabular}{l ccc c c c c}
\toprule
Operating point & Drift & $\drift_z$ & si-MSE & sens.\ (CV of $e_\ell$) & $f_{\mathrm{dep}}$ [CI] & $\rho(\drift_z,\mathrm{sens})$ [CI] & $\Delta E/E_{\mathrm{within}}$ [CI] \\
\midrule
\ca{} v2 & \TcaMITDrift & \TcaMITCm & \TcaMITSim & \TcaMITSens & \TcaMITFdep{} [\TcaMITFdepLo, \TcaMITFdepHi] & \TcaMITAone{} [\TcaMITAoneLo, \TcaMITAoneHi] & \TcaMITBone{} [\TcaMITBoneLo, \TcaMITBoneHi] \\
\ca{} 2023 & \TcaoMITDrift & \TcaoMITCm & \TcaoMITSim & \TcaoMITSens & \TcaoMITFdep{} [\TcaoMITFdepLo, \TcaoMITFdepHi] & \TcaoMITAone{} [\TcaoMITAoneLo, \TcaoMITAoneHi] & \TcaoMITBone{} [\TcaoMITBoneLo, \TcaoMITBoneHi] \\
NIID-Net & \TniidMITDrift & \TniidMITCm & \TniidMITSim & \TniidMITSens & \TniidMITFdep{} [\TniidMITFdepLo, \TniidMITFdepHi] & \TniidMITAone{} [\TniidMITAoneLo, \TniidMITAoneHi] & \TniidMITBone{} [\TniidMITBoneLo, \TniidMITBoneHi] \\
Kocsis et al. & \TkoMITDrift & \TkoMITCm & \TkoMITSim & \TkoMITSens & \TkoMITFdep{} [\TkoMITFdepLo, \TkoMITFdepHi] & \TkoMITAone{} [\TkoMITAoneLo, \TkoMITAoneHi] & \TkoMITBone{} [\TkoMITBoneLo, \TkoMITBoneHi] \\
RGB$\to$X & \TrgbxMITDrift & \TrgbxMITCm & \TrgbxMITSim & \TrgbxMITSens & \TrgbxMITFdep{} [\TrgbxMITFdepLo, \TrgbxMITFdepHi] & \TrgbxMITAone{} [\TrgbxMITAoneLo, \TrgbxMITAoneHi] & \TrgbxMITBone{} [\TrgbxMITBoneLo, \TrgbxMITBoneHi] \\
Marigold-IID & \TmgMITDrift & \TmgMITCm & \TmgMITSim & \TmgMITSens & \TmgMITFdep{} [\TmgMITFdepLo, \TmgMITFdepHi] & \TmgMITAone{} [\TmgMITAoneLo, \TmgMITAoneHi] & \TmgMITBone{} [\TmgMITBoneLo, \TmgMITBoneHi] \\
Retinex (log-domain) & \TrxMITDrift & \TrxMITCm & \TrxMITSim & \TrxMITSens & \TrxMITFdep{} [\TrxMITFdepLo, \TrxMITFdepHi] & \TrxMITAone{} [\TrxMITAoneLo, \TrxMITAoneHi] & \TrxMITBone{} [\TrxMITBoneLo, \TrxMITBoneHi] \\
Identity & 1.000 & 1.000 & \TidMITSim & \TidMITSens & \TidMITFdep{} [\TidMITFdepLo, \TidMITFdepHi] & 0.00 & \TidMITBone{} [\TidMITBoneLo, \TidMITBoneHi] \\
\midrule
\ca{} + adapter $\lambda{=}1$ & \TadMITDrift & \TadMITCm & \TadMITSim & \TadMITSens & \TadMITFdep{} [\TadMITFdepLo, \TadMITFdepHi] & \TadMITAone{} [\TadMITAoneLo, \TadMITAoneHi] & \TadMITBone{} [\TadMITBoneLo, \TadMITBoneHi] \\
\ca{} + adapter, retrained 10k & \TadtenkMITDrift & \TadtenkMITCm & \TadtenkMITSim & \TadtenkMITSens & \TadtenkMITFdep{} [\TadtenkMITFdepLo, \TadtenkMITFdepHi] & \TadtenkMITAone{} [\TadtenkMITAoneLo, \TadtenkMITAoneHi] & \TadtenkMITBone{} [\TadtenkMITBoneLo, \TadtenkMITBoneHi] \\
\ca{} + adapter, retrained 20k & \TadrerunMITDrift & \TadrerunMITCm & \TadrerunMITSim & \TadrerunMITSens & \TadrerunMITFdep{} [\TadrerunMITFdepLo, \TadrerunMITFdepHi] & \TadrerunMITAone{} [\TadrerunMITAoneLo, \TadrerunMITAoneHi] & \TadrerunMITBone{} [\TadrerunMITBoneLo, \TadrerunMITBoneHi] \\
\ca{} + adapter $\lambda{=}0.3$ & \TlamaMITDrift & \TlamaMITCm & \TlamaMITSim & \TlamaMITSens & \TlamaMITFdep{} [\TlamaMITFdepLo, \TlamaMITFdepHi] & \TlamaMITAone{} [\TlamaMITAoneLo, \TlamaMITAoneHi] & \TlamaMITBone{} [\TlamaMITBoneLo, \TlamaMITBoneHi] \\
\ca{} + adapter $\lambda{=}3$ & \TlamcMITDrift & \TlamcMITCm & \TlamcMITSim & \TlamcMITSens & \TlamcMITFdep{} [\TlamcMITFdepLo, \TlamcMITFdepHi] & \TlamcMITAone{} [\TlamcMITAoneLo, \TlamcMITAoneHi] & \TlamcMITBone{} [\TlamcMITBoneLo, \TlamcMITBoneHi] \\
\ca{} + adapter, albedo-only & \TnoimgMITDrift & \TnoimgMITCm & \TnoimgMITSim & \TnoimgMITSens & \TnoimgMITFdep{} [\TnoimgMITFdepLo, \TnoimgMITFdepHi] & \TnoimgMITAone{} [\TnoimgMITAoneLo, \TnoimgMITAoneHi] & \TnoimgMITBone{} [\TnoimgMITBoneLo, \TnoimgMITBoneHi] \\
\ca{} fine-tuned $\lambda{=}1$ & \TftaMITDrift & \TftaMITCm & \TftaMITSim & \TftaMITSens & \TftaMITFdep{} [\TftaMITFdepLo, \TftaMITFdepHi] & \TftaMITAone{} [\TftaMITAoneLo, \TftaMITAoneHi] & \TftaMITBone{} [\TftaMITBoneLo, \TftaMITBoneHi] \\
\ca{} fine-tuned $\lambda{=}3$ & \TftcMITDrift & \TftcMITCm & \TftcMITSim & \TftcMITSens & \TftcMITFdep{} [\TftcMITFdepLo, \TftcMITFdepHi] & \TftcMITAone{} [\TftcMITAoneLo, \TftcMITAoneHi] & \TftcMITBone{} [\TftcMITBoneLo, \TftcMITBoneHi] \\
\ca{} fine-tuned $\lambda{=}10$ & \TftjMITDrift & \TftjMITCm & \TftjMITSim & \TftjMITSens & \TftjMITFdep{} [\TftjMITFdepLo, \TftjMITFdepHi] & \TftjMITAone{} [\TftjMITAoneLo, \TftjMITAoneHi] & \TftjMITBone{} [\TftjMITBoneLo, \TftjMITBoneHi] \\
\ca{} + Gaussian blur & \TblurMITDrift & \TblurMITCm & \TblurMITSim & \TblurMITSens & \TblurMITFdep{} [\TblurMITFdepLo, \TblurMITFdepHi] & \TblurMITAone{} [\TblurMITAoneLo, \TblurMITAoneHi] & \TblurMITBone{} [\TblurMITBoneLo, \TblurMITBoneHi] \\
\ca{} + bilateral & \TbilatMITDrift & \TbilatMITCm & \TbilatMITSim & \TbilatMITSens & \TbilatMITFdep{} [\TbilatMITFdepLo, \TbilatMITFdepHi] & \TbilatMITAone{} [\TbilatMITAoneLo, \TbilatMITAoneHi] & \TbilatMITBone{} [\TbilatMITBoneLo, \TbilatMITBoneHi] \\
NIID-Net + adapter & \TadnMITDrift & \TadnMITCm & \TadnMITSim & \TadnMITSens & \TadnMITFdep{} [\TadnMITFdepLo, \TadnMITFdepHi] & \TadnMITAone{} [\TadnMITAoneLo, \TadnMITAoneHi] & \TadnMITBone{} [\TadnMITBoneLo, \TadnMITBoneHi] \\
\bottomrule
\end{tabular}
\end{table*}

\section{Exposure policy}
Table~\ref{tab:exposure} gives, per method, the change of the median drift ratio, contrast-matched drift and ground-truth error on OpenRooms when each frame is exposed separately instead of with one shared exposure per view, computed on the pixels valid under both policies (Section~V-F of the article). Compared on each run's own mask instead of the common pixels, the per-frame policy appears to change the ground-truth error by up to \TXOwnSimBaseMax\% for a base model; this is the mask (per-frame exposure clips fewer pixels, median valid coverage \TORvalidMedPf{} against \TORvalidMed), not the model.

\section{Paired IIW contrast and WHDR of the refinements}
Table~\ref{tab:iiw_contrast} measures, on the \TiiwN{} IIW images of the WHDR subset (every 17th of the sorted test ids, 512 pixels), the spatial contrast of each refinement's albedo relative to the base \ca{} v2 albedo (the statistic of Tables~II--III of the article, paired per image) together with the WHDR of the same outputs, so that the insensitivity of WHDR to the flattening is shown on one set of images rather than inferred across datasets (Section~V-C of the article). The base and fine-tuned albedos are brought to frame size by the archived WHDR pipeline's 8-bit bilinear resize, and the adapters are applied to that base, so the WHDR means reproduce the archived values to within GPU nondeterminism (largest deviation \TiiwKnownDev); a first run with a float resize gave means 0.0005--0.0022 lower and the same paired changes within 1.7 points. Every refinement's contrast ratio is within \TiiwMLConMaxDev{} of its multilum value (Table~III of the article). Expected-range record: ranges for the eleven contrast ratios were written before the run; nine were met, and the $\lambda = 0.3$ adapter (\TiiwlamaCon) and the $\lambda = 1$ fine-tune (\TiiwftaCon) fell below ranges that had been written from memory (0.40--0.70 and 0.30--0.60) rather than from the Table~III priors (0.14 and 0.20), with which the results agree; the pre-registered rule for a ratio below 0.3 (treat as a defect first) was applied and the WHDR known-answer check confirmed the operating points. These two misses are diagnostic-range misses, not failures of the hypotheses of Section~V-G.
\begin{table*}[!t]
\centering\footnotesize
\begin{minipage}[t]{0.415\textwidth}\centering\setlength{\tabcolsep}{5pt}
\caption{Exposure policy. Change of the median drift ratio, contrast-matched drift and ground-truth error when the OpenRooms frames are exposed per frame instead of with one shared exposure per view, computed on the pixels valid under both policies (\TORn{} views, median coverage \TORinterCov).}
\label{tab:exposure}
\begin{tabular}{l ccc}
\toprule
Method & $\Delta$ Drift & $\Delta\drift_z$ & $\Delta$ si-MSE \\
\midrule
\ca{} v2 & \TcaXDrift\% & \TcaXCm\% & \TcaXSim\% \\
\ca{} 2023 & \TcaoXDrift\% & \TcaoXCm\% & \TcaoXSim\% \\
NIID-Net & \TniidXDrift\% & \TniidXCm\% & \TniidXSim\% \\
Kocsis et al. & \TkoXDrift\% & \TkoXCm\% & \TkoXSim\% \\
RGB$\to$X & \TrgbxXDrift\% & \TrgbxXCm\% & \TrgbxXSim\% \\
Marigold-IID & \TmgXDrift\% & \TmgXCm\% & \TmgXSim\% \\
Retinex (log-domain) & \TrxXDrift\% & \TrxXCm\% & \TrxXSim\% \\
Identity & $+$0.0\% & $+$0.0\% & \TXSimIdentity\% \\
\ca{} v2 + adapter $\lambda{=}1$ & \TadXDrift\% & \TadXCm\% & \TadXSim\% \\
\ca{} v2 fine-tuned $\lambda{=}1$ & \TftaXDrift\% & \TftaXCm\% & \TftaXSim\% \\
NIID-Net + adapter & \TadnXDrift\% & \TadnXCm\% & \TadnXSim\% \\
\bottomrule
\end{tabular}
\end{minipage}\hfill
\begin{minipage}[t]{0.57\textwidth}\centering\setlength{\tabcolsep}{4pt}
\caption{Paired IIW contrast and WHDR of the refinements of \ca{} v2 (\TiiwN{} images). Contrast ratio: spatial contrast of the display-encoded, median-normalized albedo relative to the base albedo of the same image; IQR: interquartile range; images $< 0.9$: share of images whose ratio is below 0.9; WHDR on the same outputs and its change against the base.}
\label{tab:iiw_contrast}
\input{tables/tab_iiw_contrast}
\end{minipage}
\end{table*}

\section{RAW-derived renders of the multi-illumination scenes}
The control of Section~IV-A re-renders the 30 test scenes from the dataset's RAW brackets with a fixed pipeline: half-size demosaicing with linear output and no automatic brightening; a white balance fixed per camera model (the library's daylight multipliers, identical for every scene and lighting, so that no per-shot white balance enters the cross-lighting variation); the brackets of each lighting merged into radiance with relative exposure proportional to $\mathrm{ISO}\cdot t / F^2$ and a box weight that excludes clipped and noise-floor values ($0.02$--$0.95$); one exposure per scene, set by mapping the 99.9th percentile of the full 25-lighting radiance stack to $1$, so that the exposure is independent of the lighting; clipping and the standard sRGB curve. The networks therefore receive their expected display-encoded input, and the only change against the dataset's own images is that the camera's scene-dependent tone pipeline is replaced by a known fixed one. The lighting order was verified per scene by correlating the renders with the dataset's images.

\section{Noise coupling of the diffusion models}
The archived Marigold-IID runs create one torch generator per run (seed 0) and advance it through every inference, so the frames of a scene receive different noise realizations, while the Kocsis et al.\ and RGB$\to$X runs re-seed before every frame (common initial noise across frames). Table~\ref{tab:noise} re-runs Marigold-IID on the four seed-diagnostic scenes with the generator re-created before every frame (``adv.' = advancing generator). The pre-registered expectation was a fall of the drift ratio by 0--30\% of its advancing value (the five-seed diagnostic puts same-image sampling variation at \MGseedOverIllum{} of the cross-illumination variation, and the two do not add in the ratio); a fall above 50\% would have triggered recomputation of Table~II under common noise. The result is inside the expected band, so the archived values are kept and the policy is documented; the identity control scores exactly one under both policies, and the advancing-generator values reproduce the archived per-scene ratios to four decimals.
\begin{table}[t]
\centering\footnotesize
\caption{Marigold-IID drift ratio $\drift$ and contrast-matched drift $\drift_z$ on the four seed-diagnostic multilum scenes under the two noise-coupling policies, and the range of $\drift$ over four further advancing-generator seeds.}
\label{tab:noise}
\input{tables/tab_noise}
\end{table}

\section{Filtering and Retinex controls}
The two filters act on the base model's linear albedo at the evaluation resolution (frames of at most 768 pixels on multilum, 640 on OpenRooms and 512 on MIT and BigTime): a Gaussian blur with $\sigma = 4$ pixels per channel, and a bilateral filter (OpenCV) with diameter $9$, range parameter $0.1$ in linear albedo units and spatial parameter $7$ pixels; the albedo is not rescaled before filtering, so the range parameter is relative to the model's own output scale. The Retinex control is $\exp[\log(I{+}\epsilon) - G_{\sigma} * \log(I{+}\epsilon)]$ per channel on the display-encoded photograph with $\sigma = 32$ pixels and $\epsilon = 10^{-4}$, a log-domain surround; the single-scale Retinex of Jobson et al.\ forms the surround on the intensity before the logarithm, so the control is labelled ``Retinex (log-domain)'' and the article's statements about it concern this operator only.

\section{BigTime selection rule}
Sequences are visited in identifier order; for each, the frames are the first 12 PNG files in file order (masks excluded), the valid region is the intersection over those frames of the dataset's per-frame validity mask and sky mask, sequences with fewer than five frames or with less than 15\% valid pixels are skipped, and the protocol stops after 30 used sequences. Of the first 41 candidates (identifiers 0002--0128), 11 were skipped for having fewer than five frames and none for coverage; every used sequence contributed 12 frames at 512 pixels. The drift ratio treats every change between frames as a lighting change; moving objects and other scene changes that survive the masks are not separated from it.

\section{Training configurations and data provenance}
Table~\ref{tab:manifest} lists, for every trained refinement of Section~IV-C of the article, the configuration stored in its checkpoint (the operating point is bound to the checkpoint by the scoring code). Table~\ref{tab:provenance} lists the training data of the six evaluated models as stated by their authors and whether those data touch the test sets used in the article; ``unknown'' means the source does not say.

\begin{table*}[!t]
\caption{Run manifest of the trained refinements: configuration read from each checkpoint (seed 0, crops of $256^2$ throughout; adapter 0.77\,M parameters, fine-tuned albedo network 105.6\,M). LR: learning rate.}
\label{tab:manifest}
\centering\footnotesize
\input{tables/tab_manifest}
\end{table*}

\begin{table*}[!t]
\caption{Training data of the evaluated models as stated by their authors, and what the sources say about overlap with the test sets of the article (``as stated'' means not listed by the authors, not independently verified absence).}
\label{tab:provenance}
\centering\footnotesize
\input{tables/tab_provenance}
\end{table*}

\section{Controls and the record}
On every view of OpenRooms the identity scores $\drift = \drift_z = 1$ to within \TORidentDev{} and the ground-truth albedo returned as the prediction scores zero drift, zero si-MSE and a relighting error of at most \TORgtRelightMax; the constant albedo gives $\Delta E = 0$ exactly. The Spearman correlation between the drift of mismatched views and the lighting sensitivity is listed in the last column of Table~\ref{tab:supp_or}. Every archived statistic was recomputed from the 16-bit albedo cache by a second code path with zero deviation; \ca{} v2 and its main adapter were run independently on two GPU hosts and their per-view drift ratios agree to \TORxhostMax. The numerator of the drift ratio and $\mathrm{rms}(\delta)/\mathrm{rms}(R)$ of Section~III-E have a per-method Spearman correlation of \TORRhoNumMin--\TORRhoNumMax{} on OpenRooms and \TMITRhoNumMin--\TMITRhoNumMax{} on MIT.

\section{OpenRooms preparation}
Of \TORscenes{} test scenes, views were sampled uniformly (seed 0). \TORskipBlack{} of the \TORnSel{} sampled views, in \TORblackScenes{} scenes, are black renders whose emitter-free median radiance is below $10^{-3}$ in at least one lighting (at most \TORblackLumMax; a normal frame has a median between $0.5$ and $2$); they are excluded by the input-validity rule. \TORskipMask{} further views have fewer than 5\% valid pixels. The ground-truth albedo files are stored gamma-encoded and were linearized with the exponent $2.2$; the exponent was verified against the dataset's own diffuse shading: $I \approx R \odot S$ holds with a median log residual of \TGTencGLo--\TGTencGHi{} under this exponent versus \TGTencLinLo--\TGTencLinHi{} under a linear reading. After the shared-exposure tone curve the median fraction of clipped emitter-free pixels is \TORclipMedA\% (light a) and \TORclipMedB\% (light b), with a 95th percentile of \TORclipPnf\% (views whose two lightings differ strongly in brightness clip heavily in the brighter frame); per-frame exposure lowers the 95th percentile to \TORclipPnfPf\%. The median valid-mask coverage is \TORvalidMed{} (shared) and \TORvalidMedPf{} (per frame); the ground-truth albedo is below $0.02$ in some channel for a median of \TORgtBlackMed\% of the pixels (95th percentile \TORgtBlackPnf\%).

\section{Additional qualitative results}
\begin{figure*}[t]
\centering
\includegraphics[width=0.50\linewidth]{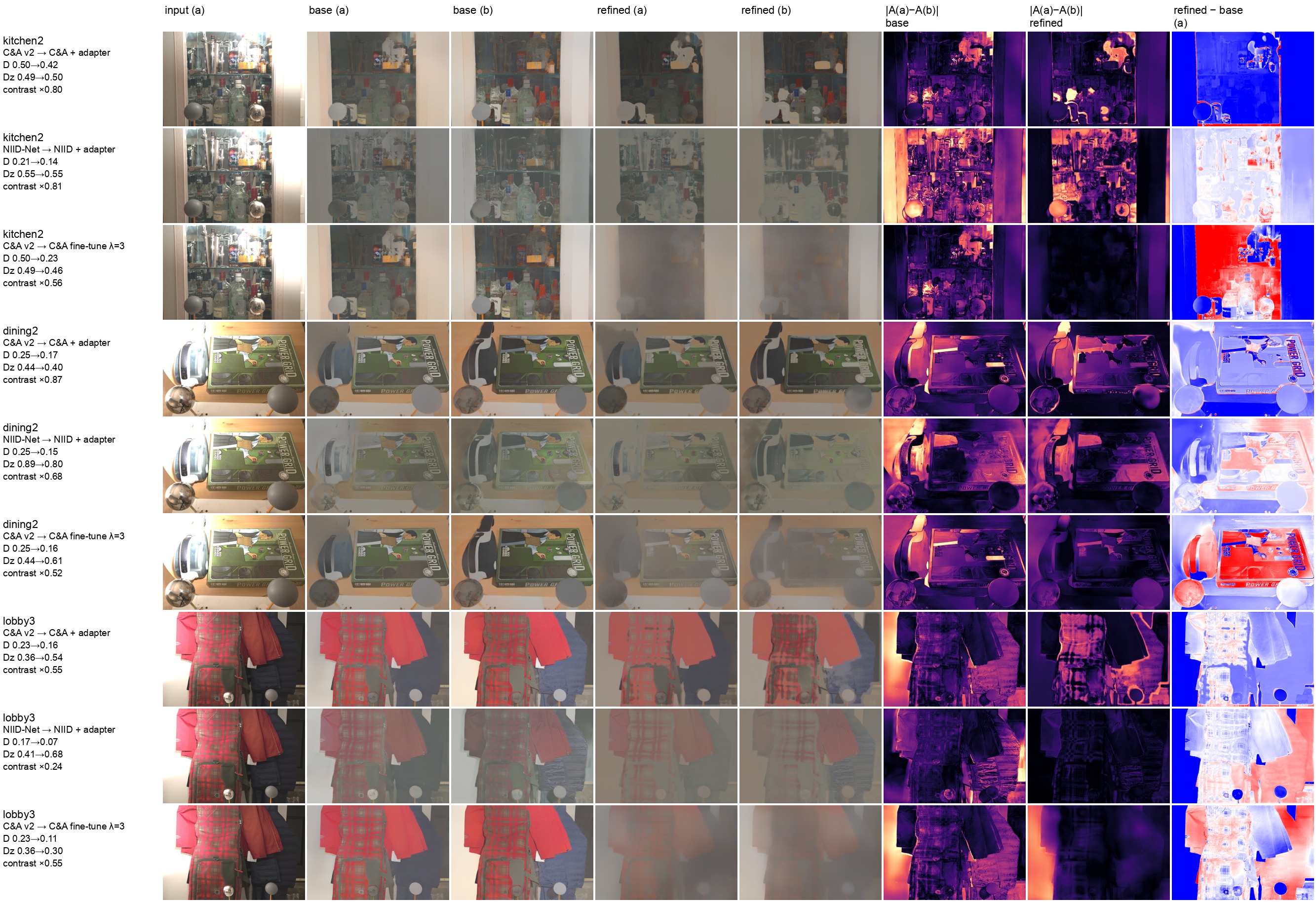}
\caption{What the consistency objective changes, for three scenes (Fig.~5 of the article shows the first): base and refined albedo under two lights, the per-pixel difference $|A_a - A_b|$ of each as a heat map (per-row range set by the base), and the signed residual (refined $-$ base) under light a on a fixed $\pm 0.25$ scale.}
\end{figure*}
\begin{figure*}[t]
\centering
\includegraphics[width=0.44\linewidth]{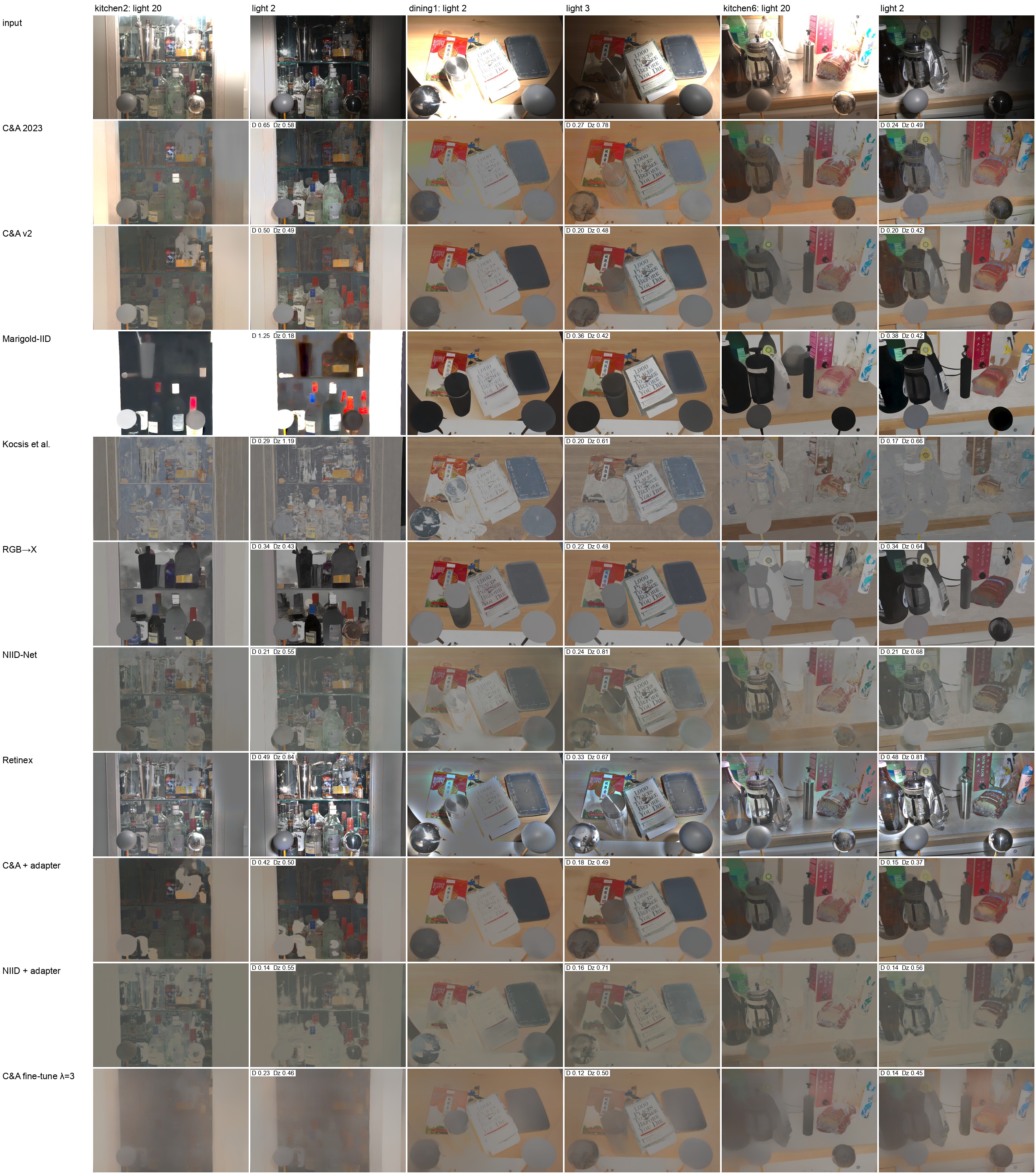}\hfill
\includegraphics[width=0.44\linewidth]{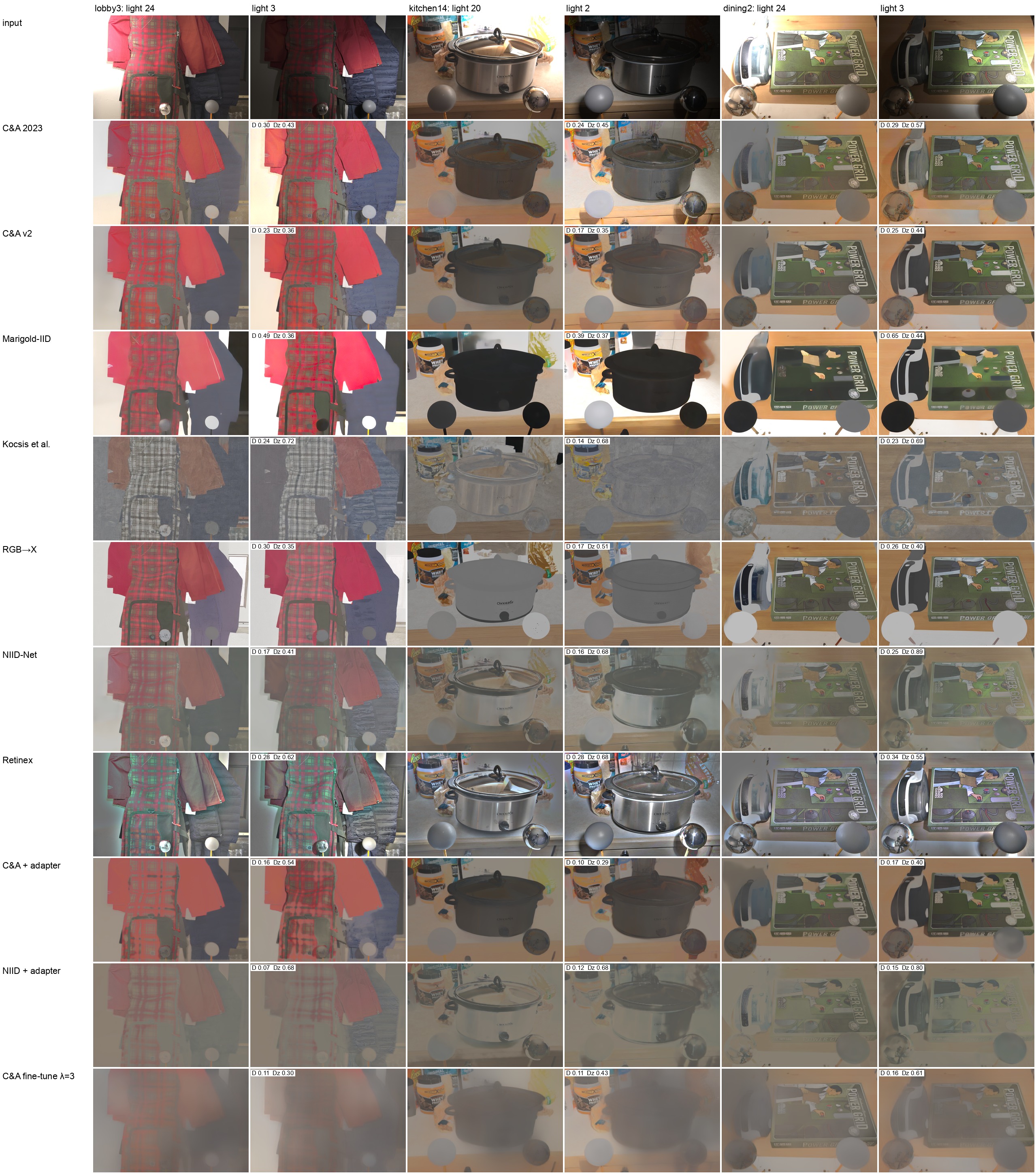}
\caption{Six multilum scenes (the two of Fig.~4 of the article and four more): input and the albedo of every method of Table~II and of three refinements under the two most different lights, with the scene's drift ratio and contrast-matched drift (computed over all 25 lightings) on the light-b panel; the two lights are the pair whose median-normalized photographs differ most.}
\end{figure*}
\begin{figure*}[t]
\centering
\begin{minipage}[t]{0.40\textwidth}\centering
\includegraphics[width=\linewidth]{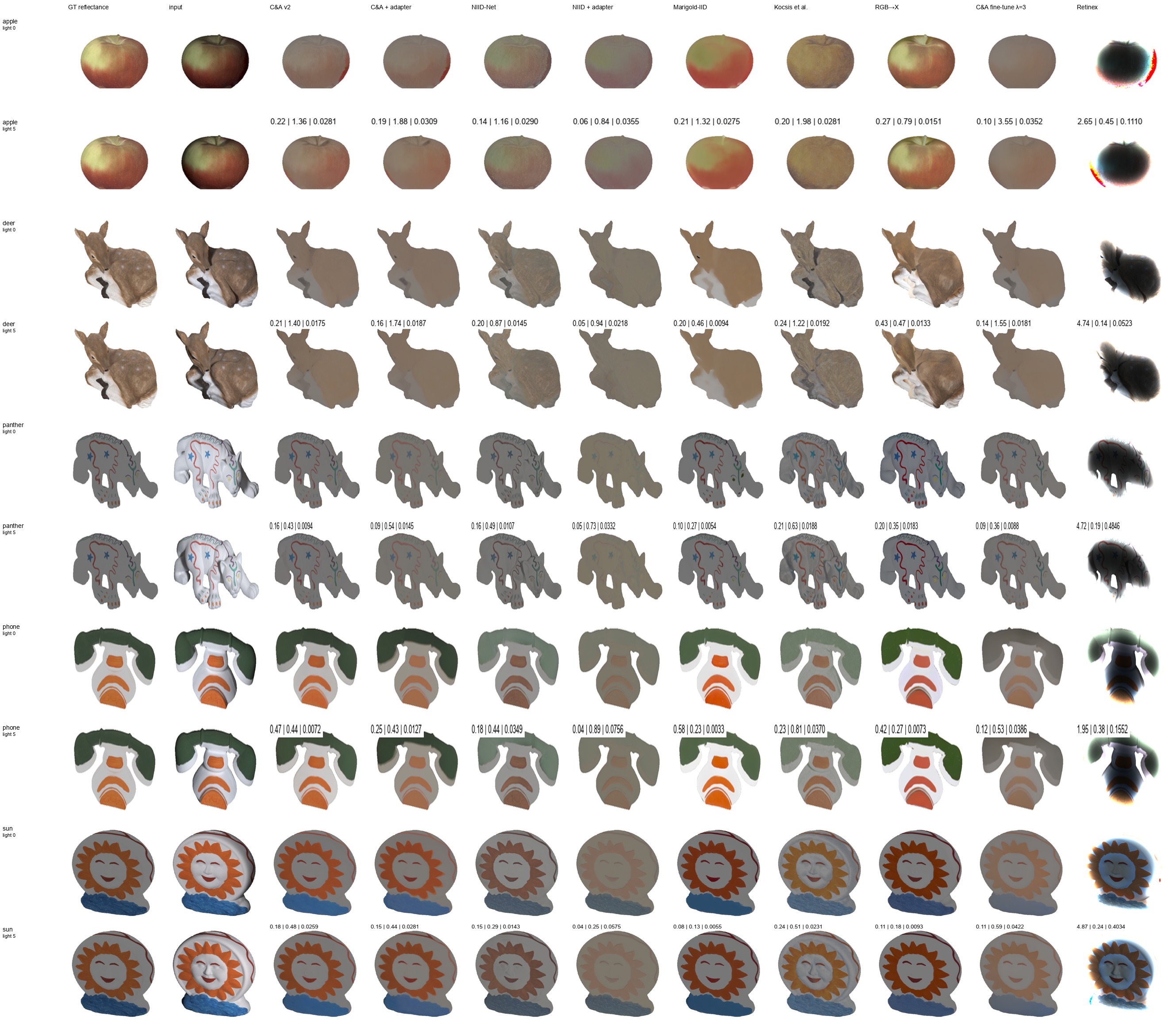}
\caption{MIT Intrinsics: five objects chosen by a fixed rule (the objects at the quintiles of \ca{} v2's drift), the ground-truth reflectance, the input under one light, and the albedo of every method with a MIT result, annotated with drift, contrast-matched drift and si-MSE.}
\end{minipage}\hfill
\begin{minipage}[t]{0.57\textwidth}\centering
\includegraphics[width=\linewidth]{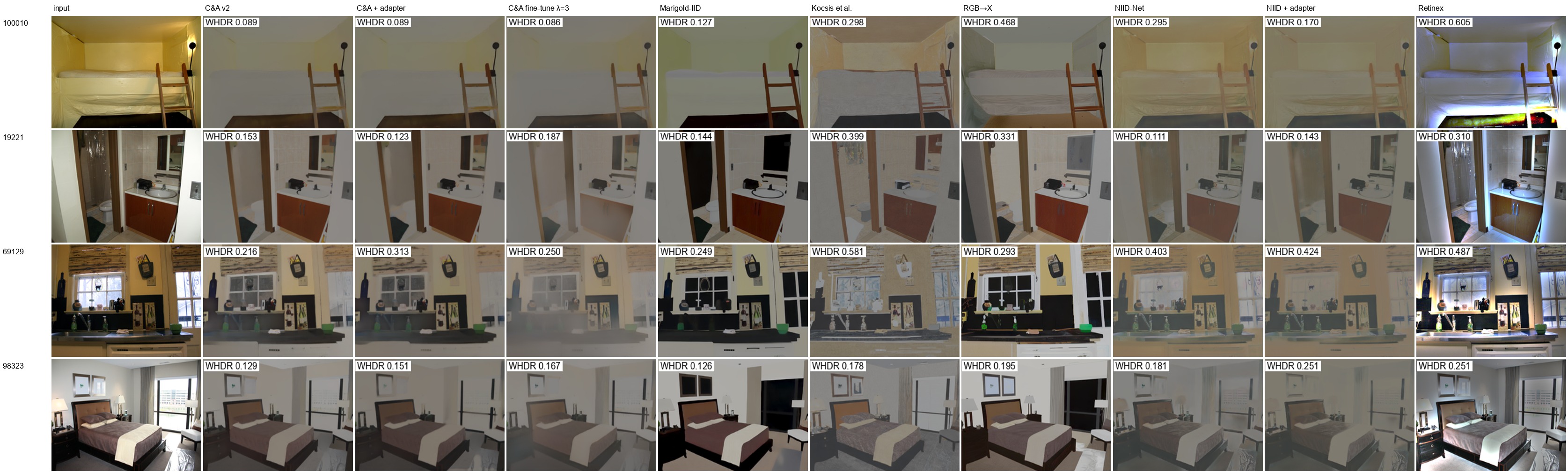}
\caption{IIW: four images chosen by a fixed rule with the albedo of every method and its per-image WHDR. The flattening of the refinements is visible and is only weakly registered by WHDR.}
\end{minipage}
\end{figure*}

\end{document}

%% file: numbers_rev.tex
\newcommand{\ADMITeSimChange}{$+$32.8\%}

\newcommand{\ADMITeSimLin}{0.0195}

\newcommand{\ADMITeSimP}{<10^{-4}}

\newcommand{\ADMLCm}{0.379}
\newcommand{\ADMLCmChange}{$-$6.9\%}
\newcommand{\ADMLCmChangeHi}{$+$0.4\%}
\newcommand{\ADMLCmChangeLo}{$-$15.6\%}

\newcommand{\ADMLCmP}{=0.230}

\newcommand{\ADMLCon}{0.41}

\newcommand{\ADMLMed}{0.143}

\newcommand{\ADMLRedMean}{30.5\%}
\newcommand{\ADMLRedMed}{28.5\%}

\newcommand{\ADNMITeSimChange}{$+$111.3\%}

\newcommand{\ADNMLCm}{0.602}

\newcommand{\ADNMLCon}{0.24}

\newcommand{\ADNMLMed}{0.119}

\newcommand{\ADNMLRedMed}{38.5\%}

\newcommand{\ADNWhdrChangeThree}{$-$10.0\%}

\newcommand{\ADNkMLCm}{0.618}

\newcommand{\ADNkMLCon}{0.24}

\newcommand{\ADNkMLMed}{0.125}

\newcommand{\ADWhdr}{0.210}

\newcommand{\ADWhdrChange}{$-$3.4\%}

\newcommand{\ADrerunMLCm}{0.381}

\newcommand{\ADrerunMLCon}{0.41}

\newcommand{\ADrerunMLMed}{0.145}

\newcommand{\ADtenkMLCm}{0.383}

\newcommand{\ADtenkMLCon}{0.40}

\newcommand{\ADtenkMLMed}{0.141}

\newcommand{\BestExistingML}{0.19}

\newcommand{\BilatMLCm}{0.404}

\newcommand{\BilatMLCon}{0.50}

\newcommand{\BilatMLMed}{0.199}

\newcommand{\BlurMLCm}{0.392}

\newcommand{\BlurMLCon}{0.48}

\newcommand{\BlurMLMed}{0.186}

\newcommand{\CABTFrac}{10\%}
\newcommand{\CABTHi}{0.719}

\newcommand{\CABTLo}{0.499}

\newcommand{\CABTMed}{0.595}

\newcommand{\CAHDRHi}{0.240}

\newcommand{\CAHDRLo}{0.203}

\newcommand{\CAHDRMed}{0.217}

\newcommand{\CAMITeSimLin}{0.0147}

\newcommand{\CAMLCm}{0.407}

\newcommand{\CAMLCon}{0.50}

\newcommand{\CAMLDif}{0.206}

\newcommand{\CAMLHi}{0.228}

\newcommand{\CAMLLo}{0.192}

\newcommand{\CAMLMed}{0.200}

\newcommand{\CAMLMin}{0.08}

\newcommand{\CAMLPc}{0.205}

\newcommand{\CAMLnAboveFifth}{15}

\newcommand{\CAWhdr}{0.217}

\newcommand{\CAoMLCm}{0.497}

\newcommand{\CAoMLCon}{0.53}

\newcommand{\CAoMLMed}{0.266}

\newcommand{\CAoWhdrThree}{0.254}

\newcommand{\FtAMLCm}{0.514}

\newcommand{\FtAMLCon}{0.10}

\newcommand{\FtAMLMed}{0.047}

\newcommand{\FtAWhdrFullChange}{$+$66.2\%}

\newcommand{\FtCMLCm}{0.454}
\newcommand{\FtCMLCmChange}{$+$11.7\%}

\newcommand{\FtCMLCon}{0.28}

\newcommand{\FtCMLMed}{0.120}

\newcommand{\FtCWhdrFullChange}{$-$0.5\%}

\newcommand{\FtJMLCm}{0.408}

\newcommand{\FtJMLCon}{0.44}

\newcommand{\FtJMLMed}{0.173}

\newcommand{\IDLINBTMed}{2.86}

\newcommand{\IDLINMLMed}{3.86}

\newcommand{\IDWhdr}{0.405}

\newcommand{\KOBTMed}{0.380}

\newcommand{\KOMLCm}{0.711}

\newcommand{\KOMLCon}{0.28}

\newcommand{\KOMLHi}{0.210}

\newcommand{\KOMLLo}{0.179}

\newcommand{\KOMLMed}{0.200}

\newcommand{\KOWhdrThree}{0.297}

\newcommand{\KOgMLDiffMed}{$+$0.004}

\newcommand{\LamAMLCm}{0.393}

\newcommand{\LamAMLCon}{0.07}

\newcommand{\LamAMLMed}{0.024}

\newcommand{\LamCMLCm}{0.411}

\newcommand{\LamCMLCon}{0.48}

\newcommand{\LamCMLMed}{0.188}

\newcommand{\MGBTMed}{0.916}

\newcommand{\MGHDRMed}{0.297}

\newcommand{\MGMLCm}{0.391}

\newcommand{\MGMLCon}{0.99}

\newcommand{\MGMLHi}{0.620}

\newcommand{\MGMLLo}{0.370}

\newcommand{\MGMLMed}{0.441}

\newcommand{\MGWhdr}{0.215}

\newcommand{\MGensMLMed}{0.408}

\newcommand{\MGseedOverIllum}{0.31}
\newcommand{\MGseedOverIllumMax}{0.44}
\newcommand{\MGseedOverIllumMin}{0.26}

\newcommand{\NIIDBTMed}{0.315}

\newcommand{\NIIDMLCm}{0.640}

\newcommand{\NIIDMLCon}{0.32}

\newcommand{\NIIDMLHi}{0.206}

\newcommand{\NIIDMLLo}{0.181}

\newcommand{\NIIDMLMed}{0.193}

\newcommand{\NIIDWhdrThree}{0.235}

\newcommand{\NoImgMLCm}{0.383}

\newcommand{\NoImgMLCon}{0.41}

\newcommand{\NoImgMLMed}{0.149}

\newcommand{\RGBXBTMed}{0.674}

\newcommand{\RGBXMLCm}{0.478}

\newcommand{\RGBXMLCon}{0.51}

\newcommand{\RGBXMLHi}{0.279}

\newcommand{\RGBXMLLo}{0.224}

\newcommand{\RGBXMLMed}{0.264}

\newcommand{\RGBXWhdrThree}{0.257}

\newcommand{\RXHDRMed}{0.285}

\newcommand{\RXMITeMed}{4.263}

\newcommand{\RXMLCm}{0.708}

\newcommand{\RXMLMed}{0.327}

\newcommand{\RXWhdr}{0.450}

%% file: numbers_tip.tex
\newcommand{\TAtwoNboot}{2,000}
\newcommand{\TAtwoSeed}{1}
\newcommand{\TAtwoSignMin}{99.85}
\newcommand{\TCmFiltORMax}{-1.3}
\newcommand{\TCmFiltORMin}{-3.9}
\newcommand{\TCmRefMLMax}{+26.3}
\newcommand{\TCmRefMLMin}{-6.9}
\newcommand{\TCmRefORMax}{+25.3}
\newcommand{\TCmRefORMin}{+0.4}
\newcommand{\TGTencGHi}{0.11}
\newcommand{\TGTencGLo}{0.04}
\newcommand{\TGTencLinHi}{0.22}
\newcommand{\TGTencLinLo}{0.09}
\newcommand{\TMITAoneBaseCiHiMax}{0.49}
\newcommand{\TMITAoneBaseMax}{0.06}
\newcommand{\TMITAoneBaseMin}{-0.52}

\newcommand{\TMITAoneRefArgmax}{the $\lambda = 1$ fine-tune}
\newcommand{\TMITAoneRefMax}{0.40}
\newcommand{\TMITAoneRefMaxHi}{0.79}
\newcommand{\TMITAoneRefMaxLo}{-0.10}
\newcommand{\TMITAoneRefMin}{-0.55}
\newcommand{\TMITAthree}{-0.90}

\newcommand{\TMITAtwoAllFive}{-0.75}

\newcommand{\TMITAtwoAllN}{18}
\newcommand{\TMITAtwoAllSix}{0.87}

\newcommand{\TMITAtwoBaseFive}{-0.26}
\newcommand{\TMITAtwoBaseFiveHi}{0.71}
\newcommand{\TMITAtwoBaseFiveLo}{-0.66}
\newcommand{\TMITAtwoBaseFiveSign}{66.9}
\newcommand{\TMITAtwoBaseSix}{0.94}
\newcommand{\TMITAtwoBaseSixHi}{1.00}
\newcommand{\TMITAtwoBaseSixLo}{0.54}

\newcommand{\TMITBoneBaseMax}{21}
\newcommand{\TMITBoneBaseMin}{5}
\newcommand{\TMITBtwoMax}{-0.19}
\newcommand{\TMITBtwoMethods}{-0.61}
\newcommand{\TMITBtwoMethodsBase}{-0.26}
\newcommand{\TMITBtwoMethodsN}{18}
\newcommand{\TMITBtwoMin}{-0.61}
\newcommand{\TMITBtwoN}{20}
\newcommand{\TMITFdepBaseMax}{28}
\newcommand{\TMITFdepBaseMin}{6}
\newcommand{\TMITRhoNumMax}{0.97}
\newcommand{\TMITRhoNumMin}{0.76}
\newcommand{\TMITaddMed}{0}
\newcommand{\TMITaddN}{20}
\newcommand{\TMITaddPnn}{1}
\newcommand{\TMITdEcrossMax}{+298.2}
\newcommand{\TMITdEcrossMin}{+0.3}

\newcommand{\TMLDriftBaseMax}{0.44}
\newcommand{\TMLDriftBaseMin}{0.19}
\newcommand{\TMixOverMatchedMax}{4.9}
\newcommand{\TMixOverMatchedMin}{2.2}
\newcommand{\TORAoneBaseCiHiMax}{0.33}
\newcommand{\TORAoneBaseMax}{0.21}
\newcommand{\TORAoneBaseMin}{-0.08}

\newcommand{\TORAoneMax}{0.28}
\newcommand{\TORAoneMin}{-0.08}

\newcommand{\TORAthree}{-0.83}
\newcommand{\TORAthreeN}{11}
\newcommand{\TORAtwoAllFive}{-0.82}

\newcommand{\TORAtwoAllN}{19}
\newcommand{\TORAtwoAllSix}{0.79}

\newcommand{\TORAtwoBaseFive}{-0.54}
\newcommand{\TORAtwoBaseFiveHi}{-0.14}
\newcommand{\TORAtwoBaseFiveLo}{-0.60}

\newcommand{\TORAtwoBaseSix}{0.66}
\newcommand{\TORAtwoBaseSixHi}{0.83}
\newcommand{\TORAtwoBaseSixLo}{0.26}

\newcommand{\TORBoneAbsMax}{0.006}
\newcommand{\TORBoneCIabsMax}{0.017}
\newcommand{\TORBoneCIexcl}{11}
\newcommand{\TORBoneMax}{0.006}
\newcommand{\TORBoneMin}{-0.003}
\newcommand{\TORBoneN}{19}
\newcommand{\TORBtwoMax}{0.21}
\newcommand{\TORBtwoMin}{-0.02}
\newcommand{\TORDriftBaseMax}{0.81}
\newcommand{\TORDriftBaseMin}{0.30}
\newcommand{\TORFdepBaseMax}{16}
\newcommand{\TORFdepBaseMin}{3}
\newcommand{\TORRhoNumMax}{0.99}
\newcommand{\TORRhoNumMin}{0.89}
\newcommand{\TORblackLumMax}{0.0006}
\newcommand{\TORblackScenes}{4}
\newcommand{\TORclipMedA}{5.0}
\newcommand{\TORclipMedB}{2.6}
\newcommand{\TORclipPnf}{59}
\newcommand{\TORclipPnfPf}{24}

\newcommand{\TORdEcrossMax}{+198.1}
\newcommand{\TORdEcrossMin}{+1.5}

\newcommand{\TORfracEcrossWorseMax}{95}
\newcommand{\TORfracEcrossWorseMin}{51}
\newcommand{\TORgtBlackMed}{11}
\newcommand{\TORgtBlackPnf}{52}
\newcommand{\TORgtRelightMax}{1.9e-14}
\newcommand{\TORidentDev}{0.0e+00}
\newcommand{\TORinterCov}{0.64}
\newcommand{\TORn}{480}

\newcommand{\TORnScenesUsed}{102}
\newcommand{\TORnSel}{500}
\newcommand{\TORscenes}{109}
\newcommand{\TORskipBlack}{14}
\newcommand{\TORskipMask}{6}
\newcommand{\TORvalidMed}{0.65}
\newcommand{\TORvalidMedPf}{0.77}

\newcommand{\TORxhostMax}{$2.3\times 10^{-3}$}

\newcommand{\TPFAtwoAllFive}{-0.70}
\newcommand{\TPFAtwoAllSix}{0.65}

\newcommand{\TXDriftBaseMax}{-3.1}
\newcommand{\TXDriftBaseMin}{-18.6}
\newcommand{\TXDriftRefMax}{-11.1}
\newcommand{\TXDriftRefMin}{-29.8}
\newcommand{\TXOwnSimBaseMax}{+14.2}

\newcommand{\TXSimBaseMax}{-1.7}
\newcommand{\TXSimBaseMin}{-4.2}
\newcommand{\TXSimIdentity}{-0.0}

\newcommand{\TadMITAone}{-0.37}
\newcommand{\TadMITAoneHi}{0.08}
\newcommand{\TadMITAoneLo}{-0.71}
\newcommand{\TadMITBone}{0.030}
\newcommand{\TadMITBoneHi}{0.046}
\newcommand{\TadMITBoneLo}{-0.015}
\newcommand{\TadMITCm}{0.551}
\newcommand{\TadMITDrift}{0.157}

\newcommand{\TadMITFdep}{0.078}
\newcommand{\TadMITFdepHi}{0.110}
\newcommand{\TadMITFdepLo}{0.044}

\newcommand{\TadMITSens}{0.121}
\newcommand{\TadMITSim}{0.0173}

\newcommand{\TadMITdEcross}{+34.1}

\newcommand{\TadORAone}{0.22}

\newcommand{\TadORAoneHi}{0.34}
\newcommand{\TadORAoneLo}{0.12}
\newcommand{\TadORBone}{-0.001}

\newcommand{\TadORBoneHi}{0.000}

\newcommand{\TadORBoneLo}{-0.003}

\newcommand{\TadORCm}{0.739}
\newcommand{\TadORCon}{0.54}

\newcommand{\TadORDrift}{0.427}

\newcommand{\TadORFdep}{0.062}
\newcommand{\TadORFdepHi}{0.079}
\newcommand{\TadORFdepLo}{0.047}

\newcommand{\TadORRp}{0.06}
\newcommand{\TadORSens}{0.162}
\newcommand{\TadORSim}{0.0144}

\newcommand{\TadORdEcross}{+32.8}
\newcommand{\TadORdSim}{+13.2}

\newcommand{\TadORfracWorse}{69}
\newcommand{\TadXCm}{-20.0}
\newcommand{\TadXDrift}{-20.5}

\newcommand{\TadXSim}{-3.4}

\newcommand{\TadnMITAone}{-0.55}
\newcommand{\TadnMITAoneHi}{-0.12}
\newcommand{\TadnMITAoneLo}{-0.81}
\newcommand{\TadnMITBone}{-0.003}
\newcommand{\TadnMITBoneHi}{0.000}
\newcommand{\TadnMITBoneLo}{-0.024}
\newcommand{\TadnMITCm}{0.852}
\newcommand{\TadnMITDrift}{0.051}

\newcommand{\TadnMITFdep}{0.006}
\newcommand{\TadnMITFdepHi}{0.012}
\newcommand{\TadnMITFdepLo}{0.003}

\newcommand{\TadnMITSens}{0.023}
\newcommand{\TadnMITSim}{0.0398}
\newcommand{\TadnMITdEcross}{+79.8}

\newcommand{\TadnORAone}{0.06}

\newcommand{\TadnORAoneHi}{0.19}
\newcommand{\TadnORAoneLo}{-0.07}
\newcommand{\TadnORBone}{-0.002}

\newcommand{\TadnORBoneHi}{-0.000}

\newcommand{\TadnORBoneLo}{-0.006}

\newcommand{\TadnORCm}{1.062}
\newcommand{\TadnORCon}{0.23}

\newcommand{\TadnORDrift}{0.226}

\newcommand{\TadnORFdep}{0.014}
\newcommand{\TadnORFdepHi}{0.019}
\newcommand{\TadnORFdepLo}{0.009}

\newcommand{\TadnORRp}{0.15}
\newcommand{\TadnORSens}{0.076}
\newcommand{\TadnORSim}{0.0245}

\newcommand{\TadnORdCm}{+7.0}

\newcommand{\TadnORdEcross}{+23.5}

\newcommand{\TadnXCm}{-3.0}
\newcommand{\TadnXDrift}{-11.1}

\newcommand{\TadnXSim}{+2.4}

\newcommand{\TadnkORAone}{0.06}

\newcommand{\TadnkORAoneHi}{0.19}
\newcommand{\TadnkORAoneLo}{-0.07}
\newcommand{\TadnkORBone}{-0.003}

\newcommand{\TadnkORBoneHi}{-0.001}

\newcommand{\TadnkORBoneLo}{-0.009}

\newcommand{\TadnkORCm}{1.032}
\newcommand{\TadnkORCon}{0.25}

\newcommand{\TadnkORDrift}{0.244}

\newcommand{\TadnkORFdep}{0.017}
\newcommand{\TadnkORFdepHi}{0.022}
\newcommand{\TadnkORFdepLo}{0.011}

\newcommand{\TadnkORRp}{0.14}
\newcommand{\TadnkORSens}{0.079}
\newcommand{\TadnkORSim}{0.0239}

\newcommand{\TadnkORdEcross}{+17.2}

\newcommand{\TadrerunMITAone}{-0.34}
\newcommand{\TadrerunMITAoneHi}{0.17}
\newcommand{\TadrerunMITAoneLo}{-0.72}
\newcommand{\TadrerunMITBone}{0.023}
\newcommand{\TadrerunMITBoneHi}{0.043}
\newcommand{\TadrerunMITBoneLo}{-0.020}
\newcommand{\TadrerunMITCm}{0.540}
\newcommand{\TadrerunMITDrift}{0.154}

\newcommand{\TadrerunMITFdep}{0.067}
\newcommand{\TadrerunMITFdepHi}{0.120}
\newcommand{\TadrerunMITFdepLo}{0.047}

\newcommand{\TadrerunMITSens}{0.129}
\newcommand{\TadrerunMITSim}{0.0165}

\newcommand{\TadrerunMITdEcross}{+25.2}

\newcommand{\TadrerunORAone}{0.18}

\newcommand{\TadrerunORAoneHi}{0.29}
\newcommand{\TadrerunORAoneLo}{0.06}
\newcommand{\TadrerunORBone}{-0.002}

\newcommand{\TadrerunORBoneHi}{-0.000}

\newcommand{\TadrerunORBoneLo}{-0.004}

\newcommand{\TadrerunORCm}{0.739}
\newcommand{\TadrerunORCon}{0.54}

\newcommand{\TadrerunORDrift}{0.415}

\newcommand{\TadrerunORFdep}{0.064}
\newcommand{\TadrerunORFdepHi}{0.080}
\newcommand{\TadrerunORFdepLo}{0.049}

\newcommand{\TadrerunORRp}{0.06}
\newcommand{\TadrerunORSens}{0.155}
\newcommand{\TadrerunORSim}{0.0142}

\newcommand{\TadrerunORdEcross}{+26.6}

\newcommand{\TadtenkMITAone}{-0.22}
\newcommand{\TadtenkMITAoneHi}{0.30}
\newcommand{\TadtenkMITAoneLo}{-0.65}
\newcommand{\TadtenkMITBone}{0.022}
\newcommand{\TadtenkMITBoneHi}{0.039}
\newcommand{\TadtenkMITBoneLo}{-0.015}
\newcommand{\TadtenkMITCm}{0.606}
\newcommand{\TadtenkMITDrift}{0.148}

\newcommand{\TadtenkMITFdep}{0.053}
\newcommand{\TadtenkMITFdepHi}{0.093}
\newcommand{\TadtenkMITFdepLo}{0.045}

\newcommand{\TadtenkMITSens}{0.128}
\newcommand{\TadtenkMITSim}{0.0179}

\newcommand{\TadtenkMITdEcross}{+65.9}

\newcommand{\TadtenkORAone}{0.18}

\newcommand{\TadtenkORAoneHi}{0.29}
\newcommand{\TadtenkORAoneLo}{0.08}
\newcommand{\TadtenkORBone}{-0.001}

\newcommand{\TadtenkORBoneHi}{-0.000}

\newcommand{\TadtenkORBoneLo}{-0.003}

\newcommand{\TadtenkORCm}{0.755}
\newcommand{\TadtenkORCon}{0.50}

\newcommand{\TadtenkORDrift}{0.387}

\newcommand{\TadtenkORFdep}{0.054}
\newcommand{\TadtenkORFdepHi}{0.069}
\newcommand{\TadtenkORFdepLo}{0.039}

\newcommand{\TadtenkORRp}{0.04}
\newcommand{\TadtenkORSens}{0.149}
\newcommand{\TadtenkORSim}{0.0144}

\newcommand{\TadtenkORdEcross}{+29.8}

\newcommand{\TbilatMITAone}{-0.32}
\newcommand{\TbilatMITAoneHi}{0.17}
\newcommand{\TbilatMITAoneLo}{-0.71}
\newcommand{\TbilatMITBone}{0.081}
\newcommand{\TbilatMITBoneHi}{0.145}
\newcommand{\TbilatMITBoneLo}{0.039}
\newcommand{\TbilatMITCm}{0.484}
\newcommand{\TbilatMITDrift}{0.195}

\newcommand{\TbilatMITFdep}{0.147}
\newcommand{\TbilatMITFdepHi}{0.278}
\newcommand{\TbilatMITFdepLo}{0.078}

\newcommand{\TbilatMITSens}{0.231}
\newcommand{\TbilatMITSim}{0.0140}

\newcommand{\TbilatMITdEcross}{+0.3}

\newcommand{\TbilatORAone}{0.18}

\newcommand{\TbilatORAoneHi}{0.29}
\newcommand{\TbilatORAoneLo}{0.06}
\newcommand{\TbilatORBone}{-0.001}

\newcommand{\TbilatORBoneHi}{-0.000}

\newcommand{\TbilatORBoneLo}{-0.003}

\newcommand{\TbilatORCm}{0.727}
\newcommand{\TbilatORCon}{0.58}

\newcommand{\TbilatORDrift}{0.463}

\newcommand{\TbilatORFdep}{0.080}
\newcommand{\TbilatORFdepHi}{0.103}
\newcommand{\TbilatORFdepLo}{0.062}

\newcommand{\TbilatORSens}{0.227}
\newcommand{\TbilatORSim}{0.0127}

\newcommand{\TbilatORdEcross}{+1.5}

\newcommand{\TblurMITAone}{-0.02}
\newcommand{\TblurMITAoneHi}{0.46}
\newcommand{\TblurMITAoneLo}{-0.51}
\newcommand{\TblurMITBone}{0.032}
\newcommand{\TblurMITBoneHi}{0.078}
\newcommand{\TblurMITBoneLo}{0.010}
\newcommand{\TblurMITCm}{0.501}
\newcommand{\TblurMITDrift}{0.161}

\newcommand{\TblurMITFdep}{0.072}
\newcommand{\TblurMITFdepHi}{0.142}
\newcommand{\TblurMITFdepLo}{0.056}

\newcommand{\TblurMITSens}{0.194}
\newcommand{\TblurMITSim}{0.0221}

\newcommand{\TblurMITdEcross}{+40.2}

\newcommand{\TblurORAone}{0.20}

\newcommand{\TblurORAoneHi}{0.30}
\newcommand{\TblurORAoneLo}{0.09}
\newcommand{\TblurORBone}{-0.000}

\newcommand{\TblurORBoneHi}{0.001}

\newcommand{\TblurORBoneLo}{-0.001}

\newcommand{\TblurORCm}{0.707}
\newcommand{\TblurORCon}{0.55}

\newcommand{\TblurORDrift}{0.436}

\newcommand{\TblurORFdep}{0.063}
\newcommand{\TblurORFdepHi}{0.082}
\newcommand{\TblurORFdepLo}{0.050}

\newcommand{\TblurORSens}{0.200}
\newcommand{\TblurORSim}{0.0143}

\newcommand{\TblurORdEcross}{+20.5}

\newcommand{\TcaMITAone}{-0.32}
\newcommand{\TcaMITAoneHi}{0.19}
\newcommand{\TcaMITAoneLo}{-0.71}
\newcommand{\TcaMITBone}{0.083}
\newcommand{\TcaMITBoneHi}{0.149}
\newcommand{\TcaMITBoneLo}{0.041}
\newcommand{\TcaMITCm}{0.482}
\newcommand{\TcaMITDrift}{0.197}

\newcommand{\TcaMITFdep}{0.153}
\newcommand{\TcaMITFdepHi}{0.281}
\newcommand{\TcaMITFdepLo}{0.079}

\newcommand{\TcaMITSens}{0.231}
\newcommand{\TcaMITSim}{0.0141}
\newcommand{\TcaORAone}{0.16}

\newcommand{\TcaORAoneHi}{0.27}
\newcommand{\TcaORAoneLo}{0.05}
\newcommand{\TcaORBone}{-0.001}
\newcommand{\TcaORBoneF}{-0.0012}
\newcommand{\TcaORBoneHi}{-0.000}
\newcommand{\TcaORBoneHiF}{-0.0001}
\newcommand{\TcaORBoneLo}{-0.004}
\newcommand{\TcaORBoneLoF}{-0.0037}
\newcommand{\TcaORCm}{0.736}
\newcommand{\TcaORCon}{0.58}
\newcommand{\TcaORDrift}{0.470}

\newcommand{\TcaORFdep}{0.083}
\newcommand{\TcaORFdepHi}{0.103}
\newcommand{\TcaORFdepLo}{0.063}

\newcommand{\TcaORRp}{0.02}
\newcommand{\TcaORSens}{0.228}
\newcommand{\TcaORSim}{0.0126}

\newcommand{\TcaXCm}{-20.7}
\newcommand{\TcaXDrift}{-18.6}
\newcommand{\TcaXDriftPf}{0.407}
\newcommand{\TcaXDriftShared}{0.499}

\newcommand{\TcaXSim}{-2.3}

\newcommand{\TcaoMITAone}{-0.52}
\newcommand{\TcaoMITAoneHi}{-0.08}
\newcommand{\TcaoMITAoneLo}{-0.79}
\newcommand{\TcaoMITBone}{0.144}
\newcommand{\TcaoMITBoneHi}{0.302}
\newcommand{\TcaoMITBoneLo}{0.030}
\newcommand{\TcaoMITCm}{0.654}
\newcommand{\TcaoMITDrift}{0.248}

\newcommand{\TcaoMITFdep}{0.172}
\newcommand{\TcaoMITFdepHi}{0.334}
\newcommand{\TcaoMITFdepLo}{0.107}

\newcommand{\TcaoMITSens}{0.283}
\newcommand{\TcaoMITSim}{0.0197}
\newcommand{\TcaoORAone}{-0.08}

\newcommand{\TcaoORAoneHi}{0.06}
\newcommand{\TcaoORAoneLo}{-0.21}
\newcommand{\TcaoORBone}{-0.000}

\newcommand{\TcaoORBoneHi}{0.016}

\newcommand{\TcaoORBoneLo}{-0.009}

\newcommand{\TcaoORCm}{1.081}
\newcommand{\TcaoORCon}{0.69}
\newcommand{\TcaoORDrift}{0.811}

\newcommand{\TcaoORFdep}{0.164}
\newcommand{\TcaoORFdepHi}{0.194}
\newcommand{\TcaoORFdepLo}{0.135}

\newcommand{\TcaoORRp}{0.00}
\newcommand{\TcaoORSens}{0.288}
\newcommand{\TcaoORSim}{0.0164}

\newcommand{\TcaoXCm}{-5.9}
\newcommand{\TcaoXDrift}{-3.1}

\newcommand{\TcaoXSim}{-2.7}

\newcommand{\TftaMITAone}{0.40}
\newcommand{\TftaMITAoneHi}{0.79}
\newcommand{\TftaMITAoneLo}{-0.10}
\newcommand{\TftaMITBone}{-0.001}
\newcommand{\TftaMITBoneHi}{0.002}
\newcommand{\TftaMITBoneLo}{-0.003}
\newcommand{\TftaMITCm}{1.743}
\newcommand{\TftaMITDrift}{0.027}

\newcommand{\TftaMITFdep}{0.001}
\newcommand{\TftaMITFdepHi}{0.003}
\newcommand{\TftaMITFdepLo}{0.001}

\newcommand{\TftaMITSens}{0.018}
\newcommand{\TftaMITSim}{0.0396}

\newcommand{\TftaMITdEcross}{+298.2}

\newcommand{\TftaORAone}{0.28}

\newcommand{\TftaORAoneHi}{0.38}
\newcommand{\TftaORAoneLo}{0.19}
\newcommand{\TftaORBone}{0.001}

\newcommand{\TftaORBoneHi}{0.002}

\newcommand{\TftaORBoneLo}{0.000}

\newcommand{\TftaORCm}{0.922}
\newcommand{\TftaORCon}{0.16}

\newcommand{\TftaORDrift}{0.119}

\newcommand{\TftaORFdep}{0.002}
\newcommand{\TftaORFdepHi}{0.003}
\newcommand{\TftaORFdepLo}{0.002}

\newcommand{\TftaORRp}{0.06}
\newcommand{\TftaORSens}{0.031}
\newcommand{\TftaORSim}{0.0327}

\newcommand{\TftaORdEcross}{+198.1}

\newcommand{\TftaXCm}{-28.4}
\newcommand{\TftaXDrift}{-29.8}

\newcommand{\TftaXSim}{-1.1}

\newcommand{\TftcMITAone}{-0.36}
\newcommand{\TftcMITAoneHi}{0.08}
\newcommand{\TftcMITAoneLo}{-0.68}
\newcommand{\TftcMITBone}{0.015}
\newcommand{\TftcMITBoneHi}{0.039}
\newcommand{\TftcMITBoneLo}{0.001}
\newcommand{\TftcMITCm}{0.584}
\newcommand{\TftcMITDrift}{0.098}

\newcommand{\TftcMITFdep}{0.024}
\newcommand{\TftcMITFdepHi}{0.050}
\newcommand{\TftcMITFdepLo}{0.017}

\newcommand{\TftcMITSens}{0.114}
\newcommand{\TftcMITSim}{0.0256}

\newcommand{\TftcMITdEcross}{+108.1}

\newcommand{\TftcORAone}{0.09}

\newcommand{\TftcORAoneHi}{0.19}
\newcommand{\TftcORAoneLo}{-0.02}
\newcommand{\TftcORBone}{0.000}

\newcommand{\TftcORBoneHi}{0.001}

\newcommand{\TftcORBoneLo}{-0.000}

\newcommand{\TftcORCm}{0.798}
\newcommand{\TftcORCon}{0.33}

\newcommand{\TftcORDrift}{0.262}

\newcommand{\TftcORFdep}{0.019}
\newcommand{\TftcORFdepHi}{0.024}
\newcommand{\TftcORFdepLo}{0.015}

\newcommand{\TftcORRp}{0.02}
\newcommand{\TftcORSens}{0.121}
\newcommand{\TftcORSim}{0.0196}

\newcommand{\TftcORdEcross}{+88.0}

\newcommand{\TftjMITAone}{-0.27}
\newcommand{\TftjMITAoneHi}{0.22}
\newcommand{\TftjMITAoneLo}{-0.69}
\newcommand{\TftjMITBone}{0.068}
\newcommand{\TftjMITBoneHi}{0.119}
\newcommand{\TftjMITBoneLo}{0.022}
\newcommand{\TftjMITCm}{0.502}
\newcommand{\TftjMITDrift}{0.194}

\newcommand{\TftjMITFdep}{0.117}
\newcommand{\TftjMITFdepHi}{0.218}
\newcommand{\TftjMITFdepLo}{0.072}

\newcommand{\TftjMITSens}{0.211}
\newcommand{\TftjMITSim}{0.0170}

\newcommand{\TftjMITdEcross}{+3.9}
\newcommand{\TftjMITdSim}{+8.8}

\newcommand{\TftjORAone}{0.17}

\newcommand{\TftjORAoneHi}{0.28}
\newcommand{\TftjORAoneLo}{0.05}
\newcommand{\TftjORBone}{-0.001}

\newcommand{\TftjORBoneHi}{0.000}

\newcommand{\TftjORBoneLo}{-0.002}

\newcommand{\TftjORCm}{0.758}
\newcommand{\TftjORCon}{0.51}

\newcommand{\TftjORDrift}{0.386}

\newcommand{\TftjORFdep}{0.060}
\newcommand{\TftjORFdepHi}{0.070}
\newcommand{\TftjORFdepLo}{0.048}

\newcommand{\TftjORRp}{-0.00}
\newcommand{\TftjORSens}{0.195}
\newcommand{\TftjORSim}{0.0139}

\newcommand{\TftjORdEcross}{+10.8}
\newcommand{\TftjORdSim}{+4.6}

\newcommand{\TgainSweepMaxDev}{10^{-7}}
\newcommand{\TgainSweepMaxUndefined}{10^{-6}}
\newcommand{\TgainSweepMinDefined}{2\times10^{-6}}
\newcommand{\TgainSweepMinGain}{10^{-8}}

\newcommand{\TidMITBone}{2.151}
\newcommand{\TidMITBoneHi}{2.423}
\newcommand{\TidMITBoneLo}{1.669}

\newcommand{\TidMITFdep}{0.387}
\newcommand{\TidMITFdepHi}{0.522}
\newcommand{\TidMITFdepLo}{0.288}

\newcommand{\TidMITSens}{0.489}
\newcommand{\TidMITSim}{0.0372}

\newcommand{\TidORBone}{0.017}

\newcommand{\TidORBoneHi}{0.029}

\newcommand{\TidORBoneLo}{0.011}

\newcommand{\TidORFdep}{0.048}
\newcommand{\TidORFdepHi}{0.067}
\newcommand{\TidORFdepLo}{0.037}

\newcommand{\TidORSens}{0.124}
\newcommand{\TidORSim}{0.0233}

\newcommand{\TidlMITFdep}{0.568}

\newcommand{\TidlMITSim}{0.0598}
\newcommand{\TidlORFdep}{0.111}

\newcommand{\TidlORSim}{0.0393}

\newcommand{\TiiwKnownDev}{0.0005}
\newcommand{\TiiwMLConMaxDev}{0.05}
\newcommand{\TiiwN}{300}
\newcommand{\TiiwadCon}{0.85}

\newcommand{\TiiwadWhdr}{+0.1}

\newcommand{\TiiwblurCon}{0.91}

\newcommand{\TiiwblurWhdr}{+22.7}

\newcommand{\TiiwftaCon}{0.15}

\newcommand{\TiiwftaWhdr}{+81.0}

\newcommand{\TiiwftcCon}{0.54}

\newcommand{\TiiwftcWhdr}{+2.4}

\newcommand{\TiiwlamaCon}{0.14}

\newcommand{\TiiwlamaWhdr}{+20.1}

\newcommand{\TkoMITAone}{0.04}
\newcommand{\TkoMITAoneHi}{0.49}
\newcommand{\TkoMITAoneLo}{-0.41}
\newcommand{\TkoMITBone}{0.051}
\newcommand{\TkoMITBoneHi}{0.115}
\newcommand{\TkoMITBoneLo}{0.022}
\newcommand{\TkoMITCm}{0.893}
\newcommand{\TkoMITDrift}{0.187}

\newcommand{\TkoMITFdep}{0.098}
\newcommand{\TkoMITFdepHi}{0.147}
\newcommand{\TkoMITFdepLo}{0.048}

\newcommand{\TkoMITSens}{0.155}
\newcommand{\TkoMITSim}{0.0253}
\newcommand{\TkoORAone}{0.12}

\newcommand{\TkoORAoneHi}{0.24}
\newcommand{\TkoORAoneLo}{0.00}
\newcommand{\TkoORBone}{-0.002}

\newcommand{\TkoORBoneHi}{-0.001}

\newcommand{\TkoORBoneLo}{-0.005}

\newcommand{\TkoORCm}{0.777}
\newcommand{\TkoORCon}{0.41}
\newcommand{\TkoORDrift}{0.301}

\newcommand{\TkoORFdep}{0.029}
\newcommand{\TkoORFdepHi}{0.037}
\newcommand{\TkoORFdepLo}{0.021}

\newcommand{\TkoORRp}{-0.11}
\newcommand{\TkoORSens}{0.130}
\newcommand{\TkoORSim}{0.0189}

\newcommand{\TkoXCm}{-15.7}
\newcommand{\TkoXDrift}{-11.1}

\newcommand{\TkoXSim}{-3.5}

\newcommand{\TlamaMITAone}{0.08}
\newcommand{\TlamaMITAoneHi}{0.51}
\newcommand{\TlamaMITAoneLo}{-0.38}
\newcommand{\TlamaMITBone}{-0.007}
\newcommand{\TlamaMITBoneHi}{-0.003}
\newcommand{\TlamaMITBoneLo}{-0.009}
\newcommand{\TlamaMITCm}{1.133}
\newcommand{\TlamaMITDrift}{0.025}

\newcommand{\TlamaMITFdep}{0.001}
\newcommand{\TlamaMITFdepHi}{0.002}
\newcommand{\TlamaMITFdepLo}{0.001}

\newcommand{\TlamaMITSens}{0.023}
\newcommand{\TlamaMITSim}{0.0362}

\newcommand{\TlamaMITdEcross}{+247.8}

\newcommand{\TlamaORAone}{0.20}

\newcommand{\TlamaORAoneHi}{0.32}
\newcommand{\TlamaORAoneLo}{0.08}
\newcommand{\TlamaORBone}{-0.000}

\newcommand{\TlamaORBoneHi}{0.000}

\newcommand{\TlamaORBoneLo}{-0.000}

\newcommand{\TlamaORCm}{0.786}
\newcommand{\TlamaORCon}{0.10}

\newcommand{\TlamaORDrift}{0.066}

\newcommand{\TlamaORFdep}{0.001}
\newcommand{\TlamaORFdepHi}{0.001}
\newcommand{\TlamaORFdepLo}{0.001}

\newcommand{\TlamaORRp}{0.11}
\newcommand{\TlamaORSens}{0.021}
\newcommand{\TlamaORSim}{0.0308}

\newcommand{\TlamaORdEcross}{+186.8}

\newcommand{\TlamcMITAone}{-0.33}
\newcommand{\TlamcMITAoneHi}{0.14}
\newcommand{\TlamcMITAoneLo}{-0.72}
\newcommand{\TlamcMITBone}{0.083}
\newcommand{\TlamcMITBoneHi}{0.145}
\newcommand{\TlamcMITBoneLo}{0.037}
\newcommand{\TlamcMITCm}{0.493}
\newcommand{\TlamcMITDrift}{0.193}

\newcommand{\TlamcMITFdep}{0.148}
\newcommand{\TlamcMITFdepHi}{0.265}
\newcommand{\TlamcMITFdepLo}{0.077}

\newcommand{\TlamcMITSens}{0.226}
\newcommand{\TlamcMITSim}{0.0145}

\newcommand{\TlamcMITdEcross}{+2.5}

\newcommand{\TlamcORAone}{0.17}

\newcommand{\TlamcORAoneHi}{0.27}
\newcommand{\TlamcORAoneLo}{0.06}
\newcommand{\TlamcORBone}{-0.001}

\newcommand{\TlamcORBoneHi}{-0.000}

\newcommand{\TlamcORBoneLo}{-0.004}

\newcommand{\TlamcORCm}{0.742}
\newcommand{\TlamcORCon}{0.57}

\newcommand{\TlamcORDrift}{0.452}

\newcommand{\TlamcORFdep}{0.078}
\newcommand{\TlamcORFdepHi}{0.101}
\newcommand{\TlamcORFdepLo}{0.060}

\newcommand{\TlamcORRp}{0.01}
\newcommand{\TlamcORSens}{0.219}
\newcommand{\TlamcORSim}{0.0128}

\newcommand{\TlamcORdEcross}{+1.9}

\newcommand{\TmgMITAone}{0.06}
\newcommand{\TmgMITAoneHi}{0.46}
\newcommand{\TmgMITAoneLo}{-0.38}
\newcommand{\TmgMITBone}{0.060}
\newcommand{\TmgMITBoneHi}{0.193}
\newcommand{\TmgMITBoneLo}{0.007}
\newcommand{\TmgMITCm}{0.365}
\newcommand{\TmgMITDrift}{0.207}

\newcommand{\TmgMITFdep}{0.280}
\newcommand{\TmgMITFdepHi}{0.354}
\newcommand{\TmgMITFdepLo}{0.178}

\newcommand{\TmgMITSens}{0.230}
\newcommand{\TmgMITSim}{0.0059}
\newcommand{\TmgNoiseAdvMed}{0.522}
\newcommand{\TmgNoiseCmRelMed}{-6}
\newcommand{\TmgNoiseFixMed}{0.480}

\newcommand{\TmgNoiseRelMax}{+2}
\newcommand{\TmgNoiseRelMed}{-7}
\newcommand{\TmgNoiseRelMin}{-10}
\newcommand{\TmgORAone}{-0.01}

\newcommand{\TmgORAoneHi}{0.13}
\newcommand{\TmgORAoneLo}{-0.14}
\newcommand{\TmgORBone}{0.006}

\newcommand{\TmgORBoneHi}{0.017}

\newcommand{\TmgORBoneLo}{0.001}

\newcommand{\TmgORCm}{0.709}
\newcommand{\TmgORCon}{0.86}
\newcommand{\TmgORDrift}{0.649}

\newcommand{\TmgORFdep}{0.107}
\newcommand{\TmgORFdepHi}{0.141}
\newcommand{\TmgORFdepLo}{0.086}

\newcommand{\TmgORRp}{0.00}
\newcommand{\TmgORSens}{0.309}
\newcommand{\TmgORSim}{0.0124}

\newcommand{\TmgXCm}{-7.6}
\newcommand{\TmgXDrift}{-13.5}

\newcommand{\TmgXSim}{-4.2}

\newcommand{\TminDen}{0.0093}
\newcommand{\TminDenZ}{0.0182}

\newcommand{\TminFrameMedian}{0.0029}
\newcommand{\TminFrameN}{53,178}
\newcommand{\TminFrameStd}{0.0041}

\newcommand{\TniidMITAone}{-0.45}
\newcommand{\TniidMITAoneHi}{0.06}
\newcommand{\TniidMITAoneLo}{-0.79}
\newcommand{\TniidMITBone}{0.068}
\newcommand{\TniidMITBoneHi}{0.135}
\newcommand{\TniidMITBoneLo}{0.006}
\newcommand{\TniidMITCm}{0.628}
\newcommand{\TniidMITDrift}{0.148}

\newcommand{\TniidMITFdep}{0.061}
\newcommand{\TniidMITFdepHi}{0.226}
\newcommand{\TniidMITFdepLo}{0.045}

\newcommand{\TniidMITSens}{0.148}
\newcommand{\TniidMITSim}{0.0182}
\newcommand{\TniidORAone}{0.21}

\newcommand{\TniidORAoneHi}{0.33}
\newcommand{\TniidORAoneLo}{0.10}
\newcommand{\TniidORBone}{-0.002}

\newcommand{\TniidORBoneHi}{-0.000}

\newcommand{\TniidORBoneLo}{-0.012}

\newcommand{\TniidORCm}{0.992}
\newcommand{\TniidORCon}{0.38}
\newcommand{\TniidORDrift}{0.376}

\newcommand{\TniidORFdep}{0.037}
\newcommand{\TniidORFdepHi}{0.046}
\newcommand{\TniidORFdepLo}{0.030}

\newcommand{\TniidORRp}{-0.01}
\newcommand{\TniidORSens}{0.143}
\newcommand{\TniidORSim}{0.0210}

\newcommand{\TniidXCm}{-11.9}
\newcommand{\TniidXDrift}{-10.0}

\newcommand{\TniidXSim}{-1.7}

\newcommand{\TnoimgMITAone}{-0.26}
\newcommand{\TnoimgMITAoneHi}{0.24}
\newcommand{\TnoimgMITAoneLo}{-0.67}
\newcommand{\TnoimgMITBone}{0.017}
\newcommand{\TnoimgMITBoneHi}{0.046}
\newcommand{\TnoimgMITBoneLo}{-0.008}
\newcommand{\TnoimgMITCm}{0.515}
\newcommand{\TnoimgMITDrift}{0.151}

\newcommand{\TnoimgMITFdep}{0.059}
\newcommand{\TnoimgMITFdepHi}{0.105}
\newcommand{\TnoimgMITFdepLo}{0.044}

\newcommand{\TnoimgMITSens}{0.131}
\newcommand{\TnoimgMITSim}{0.0179}

\newcommand{\TnoimgMITdEcross}{+48.2}

\newcommand{\TnoimgORAone}{0.20}

\newcommand{\TnoimgORAoneHi}{0.31}
\newcommand{\TnoimgORAoneLo}{0.09}
\newcommand{\TnoimgORBone}{-0.001}

\newcommand{\TnoimgORBoneHi}{0.000}

\newcommand{\TnoimgORBoneLo}{-0.002}

\newcommand{\TnoimgORCm}{0.752}
\newcommand{\TnoimgORCon}{0.52}

\newcommand{\TnoimgORDrift}{0.397}

\newcommand{\TnoimgORFdep}{0.055}
\newcommand{\TnoimgORFdepHi}{0.072}
\newcommand{\TnoimgORFdepLo}{0.042}

\newcommand{\TnoimgORRp}{0.01}
\newcommand{\TnoimgORSens}{0.174}
\newcommand{\TnoimgORSim}{0.0141}

\newcommand{\TnoimgORdEcross}{+26.7}

\newcommand{\TrgbxMITAone}{-0.22}
\newcommand{\TrgbxMITAoneHi}{0.28}
\newcommand{\TrgbxMITAoneLo}{-0.63}
\newcommand{\TrgbxMITBone}{0.211}
\newcommand{\TrgbxMITBoneHi}{0.291}
\newcommand{\TrgbxMITBoneLo}{0.141}
\newcommand{\TrgbxMITCm}{0.472}
\newcommand{\TrgbxMITDrift}{0.277}

\newcommand{\TrgbxMITFdep}{0.262}
\newcommand{\TrgbxMITFdepHi}{0.303}
\newcommand{\TrgbxMITFdepLo}{0.202}

\newcommand{\TrgbxMITSens}{0.279}
\newcommand{\TrgbxMITSim}{0.0143}
\newcommand{\TrgbxORAone}{0.12}

\newcommand{\TrgbxORAoneHi}{0.24}
\newcommand{\TrgbxORAoneLo}{-0.00}
\newcommand{\TrgbxORBone}{0.001}

\newcommand{\TrgbxORBoneHi}{0.004}

\newcommand{\TrgbxORBoneLo}{-0.000}

\newcommand{\TrgbxORCm}{0.664}
\newcommand{\TrgbxORCon}{0.71}
\newcommand{\TrgbxORDrift}{0.487}

\newcommand{\TrgbxORFdep}{0.068}
\newcommand{\TrgbxORFdepHi}{0.085}
\newcommand{\TrgbxORFdepLo}{0.055}

\newcommand{\TrgbxORRp}{0.01}
\newcommand{\TrgbxORSens}{0.212}
\newcommand{\TrgbxORSim}{0.0135}

\newcommand{\TrgbxXCm}{-11.9}
\newcommand{\TrgbxXDrift}{-9.3}

\newcommand{\TrgbxXSim}{-3.4}

\newcommand{\TrxMITAone}{0.33}
\newcommand{\TrxMITAoneHi}{0.73}
\newcommand{\TrxMITAoneLo}{-0.17}
\newcommand{\TrxMITBone}{0.002}
\newcommand{\TrxMITBoneHi}{0.003}
\newcommand{\TrxMITBoneLo}{0.001}
\newcommand{\TrxMITCm}{0.207}
\newcommand{\TrxMITDrift}{4.263}

\newcommand{\TrxMITFdep}{0.017}
\newcommand{\TrxMITFdepHi}{0.020}
\newcommand{\TrxMITFdepLo}{0.013}

\newcommand{\TrxMITSens}{0.020}
\newcommand{\TrxMITSim}{0.1685}
\newcommand{\TrxORAone}{-0.08}

\newcommand{\TrxORAoneHi}{0.14}
\newcommand{\TrxORAoneLo}{-0.29}
\newcommand{\TrxORBone}{-0.006}

\newcommand{\TrxORBoneHi}{-0.001}

\newcommand{\TrxORBoneLo}{-0.017}

\newcommand{\TrxORCm}{0.834}
\newcommand{\TrxORCon}{0.76}
\newcommand{\TrxORDrift}{0.663}

\newcommand{\TrxORFdep}{0.015}
\newcommand{\TrxORFdepHi}{0.022}
\newcommand{\TrxORFdepLo}{0.009}

\newcommand{\TrxORSens}{0.037}
\newcommand{\TrxORSim}{0.0418}

\newcommand{\TrxXCm}{-2.4}
\newcommand{\TrxXDrift}{-1.0}

\newcommand{\TrxXSim}{+0.5}

\newcommand{\TrxlMITSim}{0.2139}

\newcommand{\TrxlORSim}{0.0609}

%% file: figs/fig_refinements_key.tex
\newcommand{\RefKey}{1: adapter $\lambda{=}1$; 2: adapter $\lambda{=}0.3$; 3: adapter $\lambda{=}3$; 4: albedo-only; 5: retrain 10k; 6: retrain 20k; 7: fine-tune $\lambda{=}1$; 8: fine-tune $\lambda{=}3$; 9: fine-tune $\lambda{=}10$; 10: blur; 11: bilateral}

%% file: sections/related.tex
\section{Related Work}
\label{sec:related}
We organize prior work around one question: what has the field done with the property that defines albedo? It was stated as a definition and built into estimators as an assumption (\ref{sec:rel_def}); it was the identifying information whenever several lightings of one scene were available (\ref{sec:rel_multi}); and it is absent from the benchmarks on which methods are compared (\ref{sec:rel_eval}). Section~\ref{sec:rel_problems} names the five resulting problems, Section~\ref{sec:rel_adjacent} how adjacent fields handled them, and Table~\ref{tab:gaps} collects the requirements Section~\ref{sec:method} meets.

\subsection{Albedo as an invariant: from definition to assumption}
\label{sec:rel_def}
Barrow and Tenenbaum defined intrinsic images as surface properties that humans recover ``regardless of viewing conditions,'' recoverable only through constraints because one intensity confounds reflectance and illumination \cite{barrow78intrinsic}. Retinex theory supplied the first constraint: lightness made ``independent of flux'' by comparing ratios across edges within each spectral band rather than measuring flux \cite{land71retinex}. From then on, invariance entered the literature as the \emph{mechanism} of an algorithm---the derivative threshold separating reflectance edges from shading, the classifier labelling each derivative given a lighting direction \cite{tappen05}, physics-based losses steering a network \cite{baslamisli18retinex}, photometrically invariant edges guiding it \cite{das22pienet}---and not as a check on its output. The latest survey defines reflectance as ``the albedo invariant color of the material'' and reviews metrics that do not test the invariance \cite{garces22survey}; an earlier review grades methods by their priors and by editing under the capture light \cite{bonneel17}.

Two facts fixed by this literature bind any measurement: the recoverable quantity is defined only up to scale---a bright light on a dark surface produces the same spectrum as a dim light on a light one \cite{finlayson01pixel}; for albedo the ambiguity is a global (per-channel) gain---and the invariance is a property of a \emph{map}, the whole albedo image, not of a single number, which separates it from the illuminant estimation of Section~\ref{sec:rel_adjacent}.

\subsection{Several lightings as the identifying information}
\label{sec:rel_multi}
Whenever more than one lighting of a scene was available, the literature used the agreement of reflectance across lightings as the identifying constraint, with priors selecting among the solutions it leaves open. Weiss posed exactly this problem---$T$ images with constant reflectance and changing illumination, one reflectance to recover---and solved it by maximum likelihood with a sparse prior on illumination \cite{weiss01}; a relaxation to time-varying reflectance normalized surveillance video \cite{matsushita04tdii}; the same structure factored time-lapse video for relighting \cite{sunkavalli07}, made reflectance coherent across photo collections \cite{laffont12coherent} and decomposed static-camera sequences from temporal colour variation \cite{laffont15sequences}. Intrinsic \emph{video} made frame-to-frame coherence of the reflectance a design requirement, noting that single-image algorithms applied frame by frame are temporally incoherent \cite{ye14intrinsicvideo,bonneel14interactive,meka16live}; there the camera moves, lighting is not controlled, and coherence is built into one estimator rather than measured on others. The learned successors moved the constraint into the loss: Siamese training on illumination-varying pairs, justified by ``by definition, albedo is invariant to lighting conditions'' \cite{lettry18}; sequence losses on time-lapses \cite{li18bigtime}; a term forcing the same reflectance for two real photographs of one scene \cite{bi18hybrid}; pairs needed only at training time \cite{ma18without}; cross-projected consistency of ``photometric invariants'' across views \cite{yu19irn}; and diffusion models built for consistency across any number of views and illuminations \cite{li24idarb}. The data exist at scale: over a thousand real scenes under 25 lightings, introduced because single-illumination datasets ``do not contain the necessary supervisory signals'' \cite{murmann19multilum}.

Across these heterogeneous methods one feature recurs: agreement across lightings is \emph{consumed}, as estimator input or as training loss, and where the result is a single-image network \cite{lettry18,li18bigtime,bi18hybrid,ma18without} it is then graded on the single-light benchmarks of the next section (\cite{li18bigtime} reports MIT, IIW and SAW); sequence and multi-view estimators are graded on the fidelity of their outputs. To our knowledge no paper in this line reports, as an evaluation of the trained single-image network, how invariant its output is on held-out lightings, and none applies the constraint as a score to methods not trained with it. The measurement made here is therefore not new physics; it is the field's own identifying constraint moved from training to evaluation and applied to every method alike.

\subsection{What the evaluation protocols measure}
\label{sec:rel_eval}
Three kinds of ground truth exist, and the standard evaluations score individual predictions against them. (i) \emph{Measured decompositions}: the MIT set separates an image of each object into Lambertian shading, reflectance and specularities and scores algorithms against it \cite{grosse09mit}; the set also contains each object under ten directional lights (used in Section~\ref{sec:setup}), but the score compares one prediction with one ground truth. (ii) \emph{Human judgments within one image}: IIW crowdsourced millions of relative material comparisons at point pairs, judged ``despite variations in illumination'' inside the photograph, and scores the weighted disagreement with them (WHDR) \cite{bell14iiw}; SAW added shading judgments on such images \cite{kovacs17saw}; the ordinal data also became training targets \cite{zhou15priors,zoran15ordinal}. (iii) \emph{Rendered ground truth}: objects under HDR environment maps \cite{shi17shapenet} and physically based indoor scenes \cite{li18cgi,sengupta19nir,li20inverse,li21openrooms,roberts21hypersim,zhu22iv}, each image paired with its own albedo. The first learned decomposers were trained and graded on (iii) and (i) \cite{narihira15}; a network trained only on renders led IIW and SAW \cite{li18cgi}; supervision is dense in narrow synthetic settings, weak on diverse real scenes \cite{fan18revisit}. The models measured here---regression networks with ordinal shading \cite{careaga23intrinsic,careaga24colorful}, a normal-guided network \cite{li20niid}, and three diffusion models that sample albedo as a conditional image \cite{kocsis24iid,ke25marigold,zeng24rgbx} (the fine-tuning protocol of \cite{ke25marigold} originates in a depth estimator \cite{ke24marigold})---are all reported with these scores, as are unsupervised methods whose constraints are entirely within-image \cite{liu20usi3d}. Two adjacent diffusion-based methods are not evaluated: IntrinsicDiffusion \cite{luo24intrinsicdiffusion}, a single-image decomposer, and IntrinsicAnything \cite{chen24ia}, which regularizes optimization-based material recovery from posed images under one unknown static lighting, a different acquisition setting.

Two results already expose a blind spot. A pixel-wise decision is a strong IIW baseline, and a joint bilateral filter that smooths the reflectance layer improves the WHDR of several published methods \cite{nestmeyer17filtering}: the score improves when spatial variation is removed, be it error or material detail. On measured albedo, WHDR ``focuses only on relative albedo values and often fails to capture overall quality,'' and methods that improve it do poorly on intensity, chromaticity and texture \cite{wu23maw}. The reason is structural: ordinal relationships ``are invariant to monotonic transformations of the data'' \cite{zoran15ordinal}, and a score built from them with a tolerance band registers a contrast change only through pairs whose ratio crosses the band.

\subsection{Five problems}
\label{sec:rel_problems}
Sections~\ref{sec:rel_def}--\ref{sec:rel_eval} together give five problems, named here because Section~\ref{sec:method} answers each by name.

\textit{The Single-Light Problem.} No benchmark places the same scene under two lights and scores an arbitrary method's response, so the defining property goes untested; the coherence that intrinsic video enforces is built into one estimator and does not control the lighting. A method whose albedo is plausible in every image and different in every one---the ``shading-reflectance leakage'' the field names as a failure mode \cite{das22pienet}---can receive a similar within-image score to one whose albedo is stable.

\textit{The Open-Contrast Problem.} The within-image scores are weakly sensitive to spatial contrast \cite{zoran15ordinal,nestmeyer17filtering,wu23maw}; an invariance score has the same weakness in a stronger form, since a flat albedo varies little under any light, so invariance measured alone can be satisfied by flattening, the shortcut an objective takes when nothing penalizes it (Section~\ref{sec:rel_adjacent}).

\textit{The Encoding Problem.} A photograph is radiance passed through exposure, a camera response and an in-camera pipeline: the response is recoverable only up to scale \cite{debevec97hdr}, differs between cameras \cite{grossberg04response} and is followed by camera-specific nonlinear colour manipulations \cite{afifi19wrong,karaimer16pipeline}; exposure is a per-image choice \cite{reinhard02tone}. Models return albedo in linear or encoded units by their authors' convention, and no protocol states in which units a prediction is compared to a photograph; a ratio of the two variations is undefined until both are fixed.

\textit{The Undecomposed-Error Problem.} Rendered datasets contain the same scene under several lightings with a shared ground-truth albedo \cite{li21openrooms,shi17shapenet}, and real multi-illumination captures exist \cite{murmann19multilum}, yet we identified no split of the ground-truth error into a part that changes with the light and a part that does not. Without the split it is unknown how much of a model's error an invariance test can see, or whether lowering the invariance failure lowers the error.

\textit{The Re-use Problem.} In the applications that motivate the decomposition, albedo is re-used under a light other than the one it was estimated under: relighting from factored time-lapse \cite{sunkavalli07} or multi-view sets \cite{duchene15multiview}, relighting by modifying the illumination of a decomposed image \cite{yu20outdoor}, indoor lighting editing from a re-rendered reflectance \cite{li22editing}, and the editing tasks on which reviews compared methods \cite{bonneel17,zhou15priors} (methods predicting the relit image directly bypass the albedo \cite{philip19relighting,griffiths22outcast}). The cost of an illumination-dependent albedo error at re-use has not been measured.

\subsection{What adjacent fields did with the same problems}
\label{sec:rel_adjacent}
\textbf{Colour constancy}, the colour-only special case of our problem, set two precedents: explicit measurement of the target quantity with disciplined statistical comparison, the illuminant being measured and the field agreeing on how to summarize the error, noting that the procedure changes the ranking \cite{hordley06reevaluation,gijsenij11}; and, closer to our problem, invariant images defined operationally across lightings---a colour target under many lights and the projection that makes the patches ``equal across illuminants'' \cite{finlayson09entropy}---with temporal stability a design goal of white balance \cite{barron17ffcc}. Two things do not transfer: the target is a 3-vector, a global per-channel gain estimated under an assumption or a learned prior---grey world \cite{buchsbaum80}, grey edge \cite{vandeweijer07edge}, a learned discriminator \cite{barron15ccc}---whereas albedo is a map whose spatial contrast is a degree of freedom; and a lighting change alters shading, not only colour. TIP's Retinex line pursued illuminant invariance through the same grey-world assumption \cite{jobson97ssr,jobson97msr}, with an output contrast set by the operator's parameters; a centre/surround operator of that family therefore serves below as the non-learned control.

\textbf{Image quality assessment} met its own version of the problem by measuring the property instead of a proxy: error visibility gave way to structural similarity, validated against subjective ratings, which tells a contrast-stretched image from a blurred one of equal error \cite{wang04ssim}. The move made here---a measurement of the property in place of a within-image proxy, anchored to an external reference where one exists---follows that precedent.

\textbf{Shortcut learning and underspecification} explain why the Open-Contrast Problem is not hypothetical: a model satisfies its objective through whatever cue is available \cite{geirhos19texture,geirhos20shortcut}, pipelines with equal held-out scores behave differently in deployment \cite{damour20underspec}, and a benchmark that becomes a target stops measuring what it was built for \cite{teney20ood}. A consistency objective applied to a trained decomposer, with a fidelity score that barely sees contrast as the acceptance test, is a textbook case; Section~\ref{sec:res_probes} shows the shortcut taken.

\begin{table}[!t]
\caption{What prior work established, what it assumed, and what follows for a measurement of albedo invariance. R1--R5 are met in Section~\ref{sec:method}.}
\label{tab:gaps}
\centering\footnotesize
\begin{tabularx}{\columnwidth}{@{}>{\raggedright\arraybackslash}p{1.55cm}>{\raggedright\arraybackslash}X>{\raggedright\arraybackslash}X@{}}
\toprule
Line of work & Established / assumed & Requirement \\
\midrule
Definition \cite{barrow78intrinsic,land71retinex,finlayson01pixel} & Albedo is the illumination-invariant layer, recoverable up to a (per-channel) scale; invariance built into estimators as an assumption & R1: compare the same scene under two lights; R2: normalize scale first \\
Multi-light estimation \cite{weiss01,laffont12coherent,li18bigtime,lettry18,bi18hybrid} & Agreement across lightings identifies reflectance; used as input or loss, not, in the reviewed literature, as a standardized score for single-image methods & R1 applied to \emph{any} method, trained with the signal or not \\
Benchmarks \cite{grosse09mit,bell14iiw,kovacs17saw,li18cgi} & Within-image fidelity under one light; weakly sensitive to contrast \cite{zoran15ordinal,nestmeyer17filtering,wu23maw} & R3: a contrast-matched form and a contrast report beside any invariance score \\
Image formation \cite{debevec97hdr,grossberg04response,karaimer16pipeline,reinhard02tone} & Photograph $=$ exposed, response-encoded radiance; units unstated in protocols & R4: fixed encoding and exposure rule; identity control scoring exactly one \\
Rendered ground truth \cite{li21openrooms,shi17shapenet} & Shared albedo under several lightings available; no decomposition of the error identified & R5: split the error into light-dependent and shared parts; relate the first to R1 \\
Re-use \cite{sunkavalli07,duchene15multiview,yu20outdoor,li22editing} & Albedo is re-used under a new light; cost unmeasured & R5$'$: measure the excess error of re-use across lightings \\
Adjacent fields \cite{gijsenij11,hordley06reevaluation,wang04ssim,geirhos20shortcut} & Measure the property, not a proxy; summarize errors robustly; expect shortcuts & Medians with intervals; invariance alone is a necessary condition \\
\bottomrule
\end{tabularx}
\end{table}

%% file: tables/tab_iiw_contrast.tex
\begin{tabular}{@{}lccccc@{}}
\toprule
Refinement of \ca{} v2 & \multicolumn{2}{c}{Contrast ratio to base} & Images $< 0.9$ & WHDR & $\Delta$WHDR \\
 & median & IQR & (\%) & & (\%) \\
\midrule
\ca{} v2 (base) & 1.00 & --- & --- & 0.2085 & --- \\
adapter, $\lambda = 1$ & 0.85 & [0.71, 0.94] & 62 & 0.2088 & +0.1 \\
adapter, $\lambda = 1$, re-run 10k & 0.84 & [0.67, 0.94] & 63 & 0.2058 & -1.3 \\
adapter, $\lambda = 1$, re-run 20k & 0.86 & [0.71, 0.95] & 60 & 0.2006 & -3.8 \\
adapter, $\lambda = 0.3$ & 0.14 & [0.11, 0.18] & 100 & 0.2503 & +20.1 \\
adapter, $\lambda = 3$ & 0.97 & [0.93, 0.98] & 19 & 0.2053 & -1.5 \\
adapter, albedo only & 0.86 & [0.73, 0.95] & 61 & 0.2134 & +2.3 \\
fine-tune, $\lambda = 1$ & 0.15 & [0.12, 0.19] & 100 & 0.3775 & +81.0 \\
fine-tune, $\lambda = 3$ & 0.54 & [0.41, 0.64] & 100 & 0.2134 & +2.4 \\
fine-tune, $\lambda = 10$ & 0.84 & [0.78, 0.89] & 78 & 0.2143 & +2.8 \\
Gaussian blur ($\sigma = 4$) & 0.91 & [0.87, 0.94] & 43 & 0.2559 & +22.7 \\
bilateral filter & 1.00 & [1.00, 1.00] & 0 & 0.2004 & -3.9 \\
\bottomrule
\end{tabular}

%% file: tables/tab_noise.tex
\setlength{\tabcolsep}{3.2pt}\begin{tabular}{@{}lcccccc@{}}
\toprule
Scene & \multicolumn{3}{c}{$\drift$} & \multicolumn{2}{c}{$\drift_z$} & $\drift$, seeds 1--4 \\
 & adv. & common & change & adv. & common & (adv.) \\
\midrule
dining1 & 0.357 & 0.332 & -7.0\% & 0.423 & 0.396 & 0.354--0.366 \\
dining2 & 0.649 & 0.606 & -6.7\% & 0.436 & 0.402 & 0.646--0.677 \\
kitchen12 & 0.394 & 0.355 & -10.1\% & 0.534 & 0.503 & 0.391--0.400 \\
kitchen2 & 1.249 & 1.269 & +1.6\% & 0.181 & 0.174 & 1.160--1.296 \\
\midrule
median & 0.522 & 0.480 & -6.9\% & 0.430 & 0.399 & \\
\bottomrule
\end{tabular}

%% file: tables/tab_manifest.tex
\begin{tabular}{@{}lllrlrl@{}}
\toprule
Operating point & Base & $\lambda$ & Steps & Batch & LR & Checkpoint (sha-256 prefix) \\
\midrule
adapter $\lambda{=}1$ (adapter) & C\&A v2 & 1 & 20,000 & 4 & $2\times 10^{-4}$ & \texttt{82ab98c36c1e} \\
adapter $\lambda{=}1$, re-run, 10k (adapter) & C\&A v2 & 1 & 20,000 (checkpoint at 10,000) & 4 & $2\times 10^{-4}$ & \texttt{6a47f21b568f} \\
adapter $\lambda{=}1$, re-run, 20k (adapter) & C\&A v2 & 1 & 20,000 & 4 & $2\times 10^{-4}$ & \texttt{0c361f949d70} \\
adapter $\lambda{=}0.3$ (adapter) & C\&A v2 & 0.3 & 10,000 & 4 & $2\times 10^{-4}$ & \texttt{b47476d588d9} \\
adapter $\lambda{=}3$ (adapter) & C\&A v2 & 3 & 10,000 & 4 & $2\times 10^{-4}$ & \texttt{649a749716ba} \\
adapter, albedo-only input (adapter) & C\&A v2 & 1 & 10,000 & 4 & $2\times 10^{-4}$ & \texttt{3b7c90a0492b} \\
fine-tune $\lambda{=}1$ (fine-tune) & C\&A v2 & 1 & 10,000 & 2 pairs & $10^{-5}$ & \texttt{21efed089db1} \\
fine-tune $\lambda{=}3$ (fine-tune) & C\&A v2 & 3 & 5,000 & 2 pairs & $10^{-5}$ & \texttt{38d6cd64276f} \\
fine-tune $\lambda{=}10$ (fine-tune) & C\&A v2 & 10 & 5,000 & 2 pairs & $10^{-5}$ & \texttt{f4dc9cf76290} \\
adapter $\lambda{=}1$ (adapter) & NIID-Net & 1 & 20,000 & 4 & $2\times 10^{-4}$ & \texttt{449ec7d1ff30} \\
adapter $\lambda{=}1$, 10k (adapter) & NIID-Net & 1 & 20,000 (checkpoint at 10,000) & 4 & $2\times 10^{-4}$ & \texttt{fff57b8c17ee} \\
\bottomrule
\end{tabular}

%% file: tables/tab_provenance.tex
\begin{tabularx}{\textwidth}{@{}>{\raggedright\arraybackslash}p{2.2cm}>{\raggedright\arraybackslash}X>{\raggedright\arraybackslash}p{1.9cm}>{\raggedright\arraybackslash}p{2.3cm}>{\raggedright\arraybackslash}p{2.4cm}>{\raggedright\arraybackslash}p{2.6cm}@{}}
\toprule
Model & Training data (as stated by the authors) & Trained on the multi-illumination dataset & Includes its 30 test scenes & OpenRooms / MIT (as stated) & IIW (as stated) \\
\midrule
C\&A v2 (2024) & 8 synthetic datasets (incl. Hypersim, InteriorVerse) + Multi-Illumination (Murmann et al.); diffuse network on Hypersim only & \textbf{yes} & \textbf{unknown} (not stated by the authors) & no (InteriorVerse, Hypersim: same kind of synthetic indoor renders) & not listed \\
C\&A 2023 & Hypersim + albedo derived from the Multi-Illumination dataset (MID-Intrinsics) & \textbf{yes} (1,000 of its scenes) & \textbf{unknown} (not stated) & no (Hypersim: same kind) & not used for training \\
NIID-Net (2020) & CGIntrinsics (synthetic), per the code release & no & no & no & \textbf{unknown} (downloaded with CGIntrinsics; use in training not stated) \\
Kocsis et al. (2024) & InteriorVerse (fine-tuning Stable Diffusion v2) & no & no & no (InteriorVerse: same kind) & no \\
Marigold-IID lighting (2025) & Hypersim training split (lighting variant) & no & no & no (Hypersim: same kind) & no \\
RGB$\to$X (2024) & InteriorVerse + Hypersim + Evermotion + ImageDecomp (real images, self-labelled) & no & no & no (InteriorVerse, Hypersim: same kind) & no \\
\bottomrule
\end{tabularx}